\documentclass[12pt]{article}
\pdfoutput=1
\usepackage[round,longnamesfirst]{natbib}
\usepackage{csquotes,amsmath, amsthm, amssymb, amsbsy, mdwlist}

\usepackage[belowskip=-5pt,aboveskip=0pt]{caption}
\usepackage[hyperfigures=true]{hyperref}
\newcounter{blind}
\newcommand{\blind}[2]{\ifthenelse{\equal{\value{blind}}{0}}{#1}{#2}}
\hypersetup{
pdfauthor = {},
pdftitle = {Network Model-Assisted Inference from Respondent-Driven Sampling Data},
pdfkeywords = {Hard-to-reach population sampling; Link-tracing; Network sampling; Social networks; Exponential random graph model}
pdfsubject = {Inference from RDS data requires specialized techniques for two
reasons.  Unlike in standard sampling designs, the sampling
process is both partially beyond the control of the researcher,
and partially implicitly defined.  Therefore, it is not generally possible to
directly compute the sampling weights necessary for traditional
design-based inference. Any likelihood-based inference
requires the modeling of the complex sampling process
often beginning with a convenience sample. We
introduce a model-assisted approach, resulting in a design-based
estimator leveraging a working model for the structure of the
population over which the sampling was conducted.
We demonstrate that the new estimator has improved performance compared to
existing estimators
and is able to adjust for the bias induced by the selection of the initial
sample.}
}

\usepackage{ifthen}
\usepackage[usenames]{color}
\newcounter{content}
\newcounter{notes}
\newcounter{techreport}

\usepackage{amssymb,amsmath,amsthm}
\usepackage{graphicx}
\usepackage{mathrsfs} 
\usepackage{amsmath}
\usepackage{amssymb}
\usepackage[all]{xy}
\usepackage{color, graphics}
\usepackage{palatino, url, multicol}
\usepackage{graphicx}
\usepackage{subfigure}
\usepackage{mathrsfs}
\usepackage{multicol}
\usepackage{tabularx}

\newcommand{\bi}{\begin{itemize}}
\newcommand{\ei}{\end{itemize}}
\definecolor{Emphcolor}{cmyk}{0,0.89,0.94,0.1}
\definecolor{Netcolor}{rgb}{.8,0,.9}
\definecolor{Diseasecolor}{rgb}{1,.8,.2}
\definecolor{Sampcolor}{rgb}{0,.9,.3}
\definecolor{Black}{rgb}{0,0,0}
\definecolor{Red}{rgb}{1,0,0}
\definecolor{Blue}{rgb}{0,0,1}
\definecolor{Gray}{gray}{.6}

\newcommand{\IE}{\mathbb{E}}

\font\elevenrm=cmr11
\newcommand{\ql}{{\elevenrm ``}}
\newcommand{\qr}{{\elevenrm "}\ }
\newcommand{\qrns}{{\elevenrm "}\kern-2pt}

\theoremstyle{plain}

\theoremstyle{definition}

\newcommand{\vm}{g(\vy, \vx)}
\newcommand{\mtilde}{\tilde{g}(\vy, \vx)}

\usepackage{graphicx}
\usepackage{float}
\usepackage{rotating}
\usepackage{natbib}
\usepackage{color}
\usepackage{hypernat}
\usepackage{pdflscape}

\usepackage{graphics}
\usepackage{subfigure}
\usepackage[all]{xy}
\usepackage{amsmath}
\usepackage{amssymb}
\usepackage[all]{xy}
\usepackage{palatino, url, multicol}
\usepackage{mathrsfs}
\usepackage{multicol}
\usepackage{enumerate}

\definecolor{Emphcolor}{rgb}{.1,.1,.5}
\definecolor{Red}{rgb}{.9,0,.1}
\definecolor{Blue}{rgb}{.1,.1,.5}

\newcommand{\vx}{{\bf x}}
\newcommand{\vy}{{\bf y}}
\newcommand{\vu}{{\bf u}}
\newcommand{\vY}{{\bf Y}}

\newcommand{\bea}{\begin{eqnarray}}
\newcommand{\eea}{\end{eqnarray}}

\usepackage{ifthen}
\newboolean{draft}
\setboolean{draft}{false}
\newcommand{\knote}[1]{\ifthenelse{\boolean{draft}}{{\bf knote:~}{\it
#1}\relax}{}}
\newcommand{\mnote}[1]{\ifthenelse{\boolean{draft}}{{\bf mnote:~}{\it
#1}\relax}{}}

\newcommand{\mvh}{\hat{\mu}_{\rm VH}}

\newcommand{\mh}{\hat{\mu}_{SS}}

\title{Network Model-Assisted Inference from Respondent-Driven Sampling Data}
\author{\blind{Krista J. Gile\footnote{Assistant Professor of Statistics,
  Department of Mathematics and Statistics,
  University of Massachusetts, Amherst,
  MA 01003-9305 (E-mail: \emph{gile@math.umass.edu}).} \and Mark S. Handcock\footnote{Professor of Statistics, Department of Statistics, University of California, Los Angeles, CA 90095-1554 (E-mail: \emph{handcock@ucla.edu}).}}{}}

\begin{document}
\maketitle
\thispagestyle{empty}
\setcounter{page}{0}

\begin{abstract}
\vspace{-.05in}
Respondent-Driven Sampling is a method to sample hard-to-reach
human populations by link-tracing over their social networks.
Beginning with a convenience sample, each person sampled is given
a small number of uniquely identified coupons to distribute to other
members of the target population, making them eligible for
enrollment in the study.  This can be an effective means to collect
large diverse samples from many populations.

Inference from such data requires specialized techniques for two
reasons.  Unlike in standard sampling designs, the sampling
process is both partially beyond the control of the researcher,
and partially implicitly defined.  Therefore, it is not generally possible to
directly compute the sampling weights necessary for traditional
design-based inference. Any likelihood-based inference
requires the modeling of the complex sampling process 
often beginning with a convenience sample. We
introduce a model-assisted approach, resulting in a design-based
estimator leveraging a working model for the structure of the
population over which sampling is conducted.

We demonstrate that the new estimator has improved performance compared to existing estimators
and is able to adjust for the bias induced by the selection of the initial sample.
We present sensitivity analyses for unknown population
sizes and the misspecification of the working network model.
We develop a bootstrap procedure to compute measures of uncertainty.
We apply the method to the estimation of HIV prevalence in a population of
injecting drug users (IDU) in the Ukraine, and show how it can be extended to include 
application-specific
information.
\end{abstract}
\vspace*{.3in}

\newcommand{\N}{\mathbb{N}}
\newcommand{\s}{\mathscr{S}}

\newcommand{\q}{Q}
\newcommand{\qdn}{Q_{\N,n}}
\newcommand{\qys}{\q^*_{k,z}(y,\s)}
\newcommand{\qysk}{\q^*_{k,z}(y^i,\s^i)}
\newcommand{\mI}{\hat{\mu}_I}
\newcommand{\hatmu}{\hat{\mu}}
\newcommand{\mush}{\hat{\mu}_{SH}}

\newcommand{\p}{{\bf p}}
\newcommand{\vp}{{\bf p}}
\newcommand{\D}{{\bf D}}
\newcommand{\X}{{\bf X}}
\newcommand{\Z}{{\bf Z}}
\newcommand{\vZ}{{\bf Z}}
\newcommand{\Y}{{\mathscr{Y}}}
\newcommand{\vd}{{\bf d}}
\newcommand{\z}{{\bf z}}
\newcommand{\vz}{{\bf z}}
\newcommand{\vs}{{\bf s}}
\newcommand{\vS}{{\bf S}}
\newcommand{\vV}{{\bf V}}
\newcommand{\vv}{{\bf v}}

\newcommand{\zs}{{\bf z_s}}
\newcommand{\ds}{{\bf d_s}} 
\newcommand{\pc}{{\bf p_c}}

\newcommand{\mhs}{\hat{\mu}_{MA}}

\advance\baselineskip by -0.3pt
\section{Introduction}

There is much interest in estimating features of hard-to-reach human populations.  Such populations are characterized by the lack of a serviceable population sampling frame.  In some settings, the target population is well-connected by a network of social relations.  {\it Link-tracing} sampling strategies such as {\it snowball sampling} \citep{goodman1961} and {\it respondent-driven sampling} (RDS) \citep{heck97} are often used to leverage those social relations to sample beyond the small subgroup available to researchers.  In these settings, subsequent samples are identified and selected based on their social ties with other members of the target population.   The statistical literature dealing with such strategies \citep{goodman1961, frank71, thompson90, thompsonfrank2000},
typically assumes an idealized setting in which the initial sample is assumed to be a probability sample from the target population.  The applied literature on the other hand \citet{trow57,targetedsampling1989}, has traditionally recognized that this is impractical, and therefore treated link-tracing samples (typically referred to as snowball samples, despite Goodman's probabilistic framing) as convenience samples for which probability-based inferential methods are unfounded.  

The work of Heckathorn and colleagues \citep{heck97, salgheck04, heck07, volzheck08} around the RDS specialization of link-tracing sampling is innovative in reducing the number of links followed per respondent, such that many waves of sampling are fostered, decreasing the dependence of the final sample on the initial convenience sample.  The second main innovation of the RDS paradigm is in the {\it respondent-driven} nature of the sampling process in which subsequent samples are selected by the passing of coupons by current sample members, thus reducing the confidentiality concerns often present in hard-to-reach marginalized populations.  While this approach does reduce the dependence of the final sample on the initial sample, it is possible for substantial bias to remain based on the initial sample of seeds, as studied in simulations by \cite{gilehan10socmeth} and illustrated empirically by \cite{johnston09seeds}.  Current estimation methods \citep{heck97,  salgheck04, heck07, volzheck08, gilejasa11}, however, do not correct for biases introduced by seed selection. 
A common feature of networked populations is that the social ties are often more likely to
occur between people that have similar attributes than those who do not,
a tendency called {\it homophily} by attributes
\citep{LazMer54,Fre96,McP:01}.
In this paper we present a novel approach and inferential frame to correct for bias introduced by seed selection and for the effects of homophily.
In particular, we treat the problem of estimation of the population proportion of a binary covariate in populations where there exists homophily on the covariate of interest, based
on a branching link-tracing sample beginning with seeds selected with bias with respect to that covariate.

There is a varied formal statistical literature on inference from link-tracing network samples.  All of this work, however, involves the assumption that the initial sample is a probability sample drawn from a well-defined sampling frame, and that subsequent sampling is {\it adaptive}, or dependent on population characteristics only through their observed portions \citep{thompsonseber1996}.  In the design-based framework, these works consider cases where sampling probabilities are known for all units in the analysis \citep{goodman1961, frank71, frank2005, thompson1992, thompson2006}.  Inference is then made without reference to any superpopulation model.  In the likelihood frame, the literature treats cases where the adaptive sampling process is {\it amenable} to the model, and therefore the modeling can be conducted without explicit treatment of the sampling process \citep{thompsonfrank2000, hangile10aoas, pattison09}.  The traditional approach to RDS, originally due to \cite{heck97}, represents an alternative to this paradigm.  The assumption of the original probability sample is replaced by an assumption of sufficient waves of sampling to adequately reduce the dependence of the sample on the original sample.  

In this paper, we concern ourselves with a case in which none of these approaches suffice.  The sampling probabilities of the units are not known, making the traditional design-based approaches inadequate.  The initial sample is not a probability sample, so the sample is not adaptive or amenable, and any likelihood inference must consider the sampling process as well as the population model.  Such a joint modeling approach has been conducted in a few works \citep{franksnijders94, felixmedinathompson04, felixmedinamonjardin2006}, but each of these requires an initial probability sample from some frame to allow for modeling of the sampling process.  And while in some cases, the waves of sampling may be sufficient to suitably reduce the dependence on the initial sample, this is often not the case \citep{gilehan10socmeth}, and we are interested in the cases when this does not hold.

We begin in Section 2 by introducing respondent-driven sampling.
In Section 3, we then present our Model-Assisted inferential
approach.  Section 4 presents a simulation study illustrating the
removal of bias introduced by the initial convenience sample.
Our application to HIV prevalence estimation among injecting drug
users in the Ukraine can be found in Section 5, and Section 6 presents a
discussion and concluding remarks.

\section{Respondent-Driven Sampling}
\subsection{Notation}

We assume the target population consists of $N$ people (nodes) with labels 
$1, \ldots, N.$
Let the $N$-vector $\z$, represent a binary nodal outcome
variable of interest.  We refer
to this variable as {\ql}infection status\qrns, such that
  \[
  \z_i = \left\{\begin{array}{rl} 0 & i \textrm{ not infected}  \\
          1 & i \textrm{ infected.}  \end{array}\right. ~~~  i \in 1\ldots N
   \]
We assume the target population is connected by a network of mutual relations  
with $N\times{N}$ adjacency matrix $\vy$: 
  \[
  \vy_{ij} = \vy_{ji} = \left\{\begin{array}{rl} 1 & i \textrm{ and } j \textrm{ connected}  \\
          0 &  i \textrm{ and } j \textrm{ not connected,}  \end{array}\right.
   \]
and that this network forms a single connected component.  
Denote by $\vd_i = \sum_j \vy_{ij}$ the nodal {\it degree}, or number of network ties or {\it alters} of node $i$.  
Let $\vd =
\{\vd_1, \ldots, \vd_{N}\}.$
Denote by $\vx_i = \sum_j \z_j \vy_{ij}$ the number of network
ties node $i$ shares with infected nodes, and let $\vx = \{\vx_1,
\ldots, \vx_{N}\}.$

\subsection{Sampling Procedure}
We consider an RDS procedure of the following form:
\begin{enumerate}
  \item[0.] An small initial sample is selected from the population members accessible to researchers, typically using a convenience mechanism. They are typically 3-12 in number. They are called the {\it seeds} and comprise {\it wave} $k=0$ of the sample. 
  \item[1.] Each member of wave $k$ is given a small number (typically 2-3) of uniquely identified coupons to distribute among their alters. 
  \item[2.] Coupon recipients returning their coupons to the study center are subsequently enrolled in the study.
A person recruited in a prior wave can not be recruited again.
The wave number of a respondent is one more than that of their recruiter.
  \item[3.] Steps (1) and (2) are repeated until the desired sample size is attained.
\end{enumerate}

This process has proved effective at recruiting large and diverse samples from many hard-to-reach populations \citep{aqcdc06}, and has been widely used.  It has been heavily used in the monitoring of disease prevalence and risk behaviors among high-risk populations such as sex workers, men who have sex with men, and injecting drug users \citep{malekinejad:2008ty}, largely in the service of the reporting requirements of UNAIDS for all countries with concentrated HIV epidemics \citep{unaids2008}.  It is also used by the US Centers for Disease Control and Prevention in the behavioral monitoring of injecting drug users in 25 large US cities \citep{lanskycdcrds07}, and has also been used in other populations such as unregulated workers \citep{bernhardt2006} and jazz musicians \citep{jazzmusicians2001}.


We represent the full sampling mechanism by the random variables:
    \[
  \vS^{k}_i = \left\{\begin{array}{rl} 
          1 & \textrm{ person\ } i \textrm{ is sampled in wave } k\\ 
          0 & \textrm{ otherwise} \end{array}\right.  ~~~  i \in 1\ldots N , k \in 0, \ldots.
   \]
  \[
  \vS_i = \sum_{k=0}^{\infty}{\vS^{k}_i}=\left\{\begin{array}{rl} 1 & \textrm{ person\ } i \textrm{ is sampled} \\
          0 & \textrm{ person\ } i \textrm{ is not sampled}  \end{array}\right.  ~~~  i \in 1\ldots N,
   \]
and let $\vs^{k}$ denote the observed sampling vector corresponding to the people sampled
in wave $k$.
Based on the sampling procedure, we exactly observe the elements of
$\vz, \vd$ and $\vx$ 
corresponding to $i: \vs_i=1$.  A variant when $\vx$ cannot be observed directly,
as in the application in Section 5, substitutes an estimate of $\vx$ based on observed referral patterns.

Further we assume each respondent distributes a number of coupons completely at random from among their alters, with the number determined by a common distribution.

\subsection{Design-based Inferential Approach}
\label{sec:designbased}
We consider design-based estimators for the population mean 
$\mu=\frac{1}{N}\sum_{i=1}^N {\vz_i}$.
Because the sampling probabilities of the people selected through RDS are almost never explicitly known, we follow \cite{volzheck08}, and \cite{gilejasa11} in constructing a model for the sampling process, and estimating sampling probabilities accordingly.   
We use a generalized Horvitz-Thompson estimator of the form:

\vskip -0.3in
\newcommand{\piihat}{\hat{\pi}_{i}}
\bea
 \hat{\mu} =  \frac{\sum_{i=1}^N \frac{\vS_i \vz_i}{\piihat}  }
 {\sum_{i=1}^N \frac{\vS_i }{\piihat}  },
 \label{ht}
\eea
where estimated sampling probabilities $\piihat = \IE(\vS_i \vert \s)$ are computed under an approximation $\s$ to the true RDS sampling process.  
If the inclusion probabilities were known this estimator is referred to as the H\"ajek estimator \citep{lumley2010}, and typically performs better than the corresponding Horvitz-Thompson estimator \citep{sarndal1992}.
The estimators introduced by \cite{volzheck08} and \cite{gilejasa11} differ, and ours further differs, in their specification of the sampling process $\s$.

Most inference from RDS data approximates the sampling process as a with-replacement random walk on the space of graph nodes, with transitions along the edges or social relations.  For the purpose of inference, sampling is treated as a Markov chain at equilibrium \citep{salgheck04, volzheck08}.  Such inference involves sampling weights proportional to the self-reported degrees which are the equilibrium sampling probabilities  of the with-replacement random walk on a connected network.  While this is a useful first approximation, it has several limitations.  First, as highlighted in \cite{gilejasa11}, this type of inference does not respect the without-replacement nature of the sampling process, which can lead to biased estimates.  \cite{gilejasa11} presents an approach correcting for this feature by substituting a without-replacement successive sampling approximation to the sampling process.  Neither this, nor earlier estimators, however, address the fundamental issue of bias induced by the selection of the initial sample.  Such bias is illustrated in \cite{gilehan10socmeth}, as well as in the current paper, and correction for it is a key contribution of the present paper.

As with these earlier approaches, the first requirement of our sampling model is that it account for the different sampling probabilities by nodal degree.  Unlike these other approaches, we further require our approach to account for the bias introduced by the selection of seeds in the presence of network homophily in the underlying population.  This requires consideration of features of the social network, $\vy$, in particular the homophily of the relations.

We make no assumptions about the mechanism for selecting the initial sample and will condition on the seed characteristics throughout the analysis.  

If the network $\vy$ were fully known, we could use simulation to estimate the sampling probability $\hat{\pi}_{i, \vy} = \IE(\vS_i \vert \s, \vy, \vs^0)$ of each node, conditional on the selection of seeds, $\vs^0$. Explicitly, we would repeatedly simulate RDS under sampling model $\s$ starting from $\vs^0$ each time and compute the 
$\hat{\pi}_{i, \vy}$ as the proportion of simulated samples containing node $i.$ 
These could be used in (\ref{ht}) to form an estimator.   Unfortunately, $\vy$ is typically only partially known, and so we apply a model-assisted approach.

\section{A Model-Assisted Approach}
\label{sec:alternatives}

Our approach is an extension of the model-assisted design-based approaches presented in \cite{sarndal1992}.  Existing work in this area uses a working model form to construct estimators that are (approximately) design-unbiased, whether the model holds or not, and have smaller design variance if the model does hold. Our case is slightly different.  The sampling process we consider is only locally defined, and originates at a sample with unknown distribution.  We therefore cannot guarantee design-unbiasedness.  In fact, we require reference to a model form to recover approximate design-unbiasedness, rather than to improve efficiency.  This is because the impact of the seed characteristics on the subsequent sample is mediated by the structure of the underlying social network.  

Our approach is to assume a working superpopulation model from which the network was drawn and use it to estimate sampling probabilities conditional on the selection of the initial sample.  

\subsection{Network Working Model}

We consider models of {\it exponential-family random graph model} (ERGM) form \citep{fra86,hunhan06,hungoodhan07}, conditional on the set of nodal degrees and infection statuses, and including a single additional parameter representing homophily on $\vz$.  In particular:

\vskip -0.3in
\bea
P(\vY=\vy \vert \vz, \vd, \eta) = \frac{\exp({\eta g(\vy, \vz)})}{c(\eta \vert \vz, \vd)},
\label{ergm}
\eea
where $g(\vy, \vz) = \sum_{i,j = 1}^N \vy_{ij} \vz_i (1-\vz_j)$, and 
$c(\eta \vert \vz, \vd) = \sum_{\vu \in \mathscr{Y}(\vz, \vd)} \exp(\eta g(\vu, \vz))$, 
and the space $\mathscr{Y}(\vz, \vd)$ consists of all binary undirected networks consistent with $\vd$ and $\vz$ (the dependence on $\vd$ and $\vz$ is suppressed below).
Note that this model form, as well as the simulation procedure to follow, requires knowledge
of the population size $N$.

Given this model form, we use the estimator (\ref{ht}) based on sampling weights assumed constant over equivalence classes by degree and infection status and estimated under the model:
\[
\hat{\pi}_{i,\eta} = \IE(\vS_i \vert \s, \vz, \vd, \eta, \vs^0) = \sum_{\vy \in \mathscr{Y}(\vz, \vd)} \hat{\pi}_{i,\vy} P(\vY=\vy \vert \vz, \vd, \eta), 
\] 
where $\hat{\pi}_{i, \vy} = \IE(\vS_i \vert \s, \vy, \vs^0)$, as defined in Section \ref{sec:designbased}.
Note that to treat these equivalence classes, we condition on the equivalence classes of the seed nodes selected, rather than the unique identities of those nodes.

We also do not know the network working model parameter $\eta$, and therefore must estimate it from the available data. The estimator is then computed using sampling probabilities based on the estimated network working model given by $\hat{\eta}$:  
\newcommand{\piihatetahat}{\hat{\pi}_{i,\hat{\eta}}}
\newcommand{\pihatetahat}{\hat{\pi}_{\hat{\eta}}}
\bea
\hat{\pi}_{i,\hat{\eta}} = \IE(\vS_i \vert \s, \vz, \vd, \hat{\eta}, \vs^0) = \sum_{\vy \in \mathscr{Y}(\vz, \vd)} \pi_{i,\vy} P(\vY=\vy \vert \vz, \vd, \hat{\eta}) 
\label{pietahat}
\eea

These are the estimated probabilities used in our proposed estimator.  This requires fitting a network working model to data sampled through RDS, which we address in the next section.

\subsection{Fitting the Network Working Model} 

\cite{thompsonfrank2000} and \cite{hangile10aoas} provide an approach to fitting models of form similar to (\ref{ergm}) to data sampled through link-tracing samples.  Unfortunately, these approaches require a sample that is {\it amenable} to the model in question.  That is:

\bea
P(\vS \vert \vy, \vd, \vz) = P(\vS \vert \vy_{obs}, \vd_{obs}, \vz_{obs}),
\label{mar}
\eea
where $*_{obs}$ represents the observed part of $*$, and also that the sampling and model parameters are separable.  This is equivalent to the conditions for {\it ignorability} according to \cite{rub76} and \cite{litrub2002}.  Unfortunately, in the case of RDS, condition (\ref{mar}) is violated by the convenience sample of seeds, which may well depend on unobserved characteristics.

Therefore, we require a novel approach to model fitting. As $\vd$ and $\vz$ are unknown, we construct design-based estimators of them from estimates
of the sampling probabilities $\piihat.$ Specifically,  let
$\N_{kl}$ be the number of nodes of degree $k$ and infection 
status $l$, $k \in \{1,\ldots, N-1\}$, $l \in \{0,1\}$ and
$\N=\{\N_{kl}\}_{k=1;l=0}^{k=N-1;l=1}.$ We estimate $\N$ and $g(\vy, \vz)$ by,
\vskip -20pt
\begin{align}
  {\tilde \N}_{kl} &=  {1\over{N}}\sum_{i=1}^{N} \frac{\vS_i \mathbb{I}({\vd_i=k, \vz_i=l})}{\piihat} \label{nhat}\\
  {\mtilde} &=  \sum_{i=1}^{N} \frac{\vS_i \left( \vx_i (1-\vz_i) + (\vd_i - \vx_i) \vz_i   \right)}{2 \piihat} \label{mhat}
\end{align}
where
$\mathbb{I}(*)$ is the indicator function on $*$, and $\hat{\pi}_{i}$ is assumed constant for all $i: \vd_i=k, \vz_i=l$.  
Note that this requires the observation of $\{\vx_i : \vS_i = 1, i=1, \ldots, N\}$.  
We then estimate $\eta$ as the natural parameter corresponding to the mean value parameter ${\mtilde}$ with the joint degree and infection status sequence
implied by ${\tilde \N}.$ Details of this computation are given in the Supplemental Materials.
\mnote{For ${\tilde \N}_{kl} $ in (5) why not divide by the $\sum_{i=1}^{N} \frac{\vS_i }{\piihat}$ instead of
$N$?}

\subsection{Algorithm}
Note that the value of the network working model parameter, required to estimate $\pi,$ in turn, depends on the value of $\pi$.  We therefore apply an approach similar to self-consistency \citep{leemeng07} to find a joint solution to (\ref{nhat}) and (\ref{mhat}), as well as to the 
equations:
\bea
\piihatetahat = \IE\left(\vS_i \vert \s, \vz, \vd, \hat{\eta}({\tilde \N}, {\mtilde})\right)~~~~~~~~~~~~~~i=1,\ldots, N.
\label{probseqn}
\eea

This approach iterates between estimating the network working model parameter given
values for the sampling probabilities, and then estimating the sampling probabilities given the network working model parameter. Explicitly, it is:

\bi
  \item Estimate $\pihatetahat$ proportional to degree $d_i$.
  \item Iterate the following steps:
  \bi
  \item Compute design-based estimates of statistics $ {\tilde \N}_{kl}$ and $ {\mtilde}$ using $\pihatetahat$ in (\ref{nhat}) and (\ref{mhat}).
  \item Determine the working ERGM parameter $\eta$ corresponding to ${\tilde \N}$ and $ {\mtilde}$. 
  \item Simulate $M$ networks according to the working ERG model.
Estimate $\pihatetahat$ by simulated RDS sampling from the resulting networks.
  \ei
  \item Use the resulting estimated probabilities, $\pihatetahat$, to form the weighted estimator of the quantity of primary interest:
\bea
 \hat{\mu}_{MA} =  \frac{\sum_{i=1}^N \frac{\vS_i \vz_i}{\pihatetahat}  }
 {\sum_{i=1}^N \frac{\vS_i }{\pihatetahat}  }.
 \label{ht2}
\label{estmu3}
\eea

\ei

The iterative nature of this procedure is similar to that used
for the successive sampling estimator of \citet{gilejasa11}.
This algorithm differs in the core process of estimating the
inclusion probabilities.  More details of this procedure are provided in the supplemental materials.

The simulation procedure implicit in this estimation algorithm lends itself to a realistic bootstrap approach to standard error estimation.  We present such a bootstrap in the supplemental materials, along with a simulation study illustrating its performance under a variety of conditions.

\section{Comparing the Model-Assisted to Existing Estimators: A Simulation Study}
\label{sec:simstudy}

\cite{gilehan10socmeth} present an extensive simulation study of RDS based, where possible,
on a set of realistic characteristics of
data from the CDC pilot study of RDS \citep{aqcdc06}.
For comparison purposes, our simulation study uses the same simulated populations as \cite{gilejasa11}, along with extensions necessary for our sensitivity analyses.

\subsection{Study Design}

Our simulation study is designed around three levels of simulation:
\bi
  \item The generation of random networks according to specified network features 
  \item The generation of simulated RDS samples from each network
  \item The estimation of infection prevalence from each set of simulated sample data.
\ei
We use variants of the network and sampling parameters to study the behavior of the proposed estimator. Descriptions and  levels of these parameters are listed in Tables \ref{tab:simparams} and \ref{tab:sampparams}.

\begin{table}[h] \caption{Parameters of simulated networks.  Default parameters given in boldface.}
\begin{center}
\begin{tabular}{l||l|c}
Parameter & Meaning & Values \\
\hline
\hline
Number of nodes & & {\bf 1000}, 715\\
\hline
Prevalence & $\mu = \frac{1}{N}\sum_i{\vz_i}$ & {\bf 0.20} \\
\hline
Mean degree & $\bar{\vd} = \frac{1}{N} \sum_{i=1}^N \IE (d_i) = \frac{1}{2} \sum_{i,j=1}^N \IE (Y_{ij})$ & {\bf 7} \\
\hline
Homophily & $R = \frac{\frac{2}{N^1(N^1-1)}\sum_{i,j} \vz_i \vz_j \IE (y_{ij}) }{\frac{1}{N^1N^0}\sum_{i,j} \vz_i (1-\vz_j) \IE (y_{ij}) }$ & {\bf 5}, 3, 1\\
&where $N^0 = N(1-\mu), N^1 = N\mu$ &\\
\hline
Activity &$w = \frac{\bar{d}^1}{\bar{d}^0} = \frac{\frac{1}{N^1}\sum_{i,j} \vz_i \IE(y_{ij})   }{\frac{1}{N^0} \sum_{i,j} (1-\vz_i) \IE(y_{ij})   }$ & {\bf 1}, 1.8\\
Ratio&&\\ [-0.3in]
\end{tabular}\label{tab:simparams}
\end{center} 
\end{table}

To allow for comparability across simulation conditions, throughout our simulations, we maintain the same true recoverable prevalence, $\mu= 0.20$, the same sample size $n=500$, and the same mean degree ${\bar{\bf d}} = 7$.  We consider variations on the population size (hence the sample fraction), the degree of clustering or {\it homophily} on infection status, and differential rates of tie formation by infection status (or {\it activity ratio}).  The parameter levels considered are summarized in Table \ref{tab:simparams}.

Under each set of network parameters, networks are simulated according to an ERGM with sufficient statistics:
\begin{align}
  \label{eqn:simmodel}
  g_1(\vy) & = \sum_{i=1}^N \sum_{j<i} \vy_{ij}\vz_i \vz_j  \notag \\
  g_2(\vy) & =  \sum_{i=1}^N \sum_{j<i} \vy_{ij}(1-\vz_i)(1- \vz_j)  \\ 
  g_3(\vy) & =  \sum_{i=1}^N \sum_{j=1}^N \vy_{ij}\vz_i(1- \vz_j). \notag 
\end{align}
These three terms correspond to the unique cells of the $2 \times 2$ mixing matrix on $\vz$, and for a given number of nodes $N$ and prevalence $\mu$, are uniquely defined by $\bar{\bf d}$, $R$, and $w$.  
Note that this model is similar to (\ref{ergm}), but not identical. While (\ref{ergm})
conditions on the fixed degree of each node, 
this model allows for stochastic variability in degrees around mean value parameters
given by (\ref{eqn:simmodel}).

From each simulated network, a single RDS sample is drawn according to parameters in Table \ref{tab:sampparams}.  A fixed number $n^0$ of seed nodes are selected with probability proportional to degree (the best case for the earlier estimators), from either the full population or from the infected nodes only (to simulate extreme seed bias).  The simulated process treats the case of two coupons distributed by each respondent completely at random among its previously un-sampled alters.  Two coupons are chosen for simplicity, and because it represents the sampling process better than either 3 (equating to the return of all coupons in practice) or 1 (resulting in non-branching chains).

\begin{table}[h]\caption{Parameters of simulated RDS sampling.  Default parameters given in boldface.}
\begin{center}
\begin{tabular}{l|l|c} 
Parameter & Meaning & Values \\
\hline
\hline
Number of Seeds & $n^0 = \sum_i \vS_i^0$ & {\bf 10}, 6, 20 \\
\hline
Seed Selection & Sequentially with probability proportional  & {\bf full population}, \\
& to degree from either: & infected nodes \\
\hline
Branching & From each sampled node, up to $n_{cup}$ & {\bf 2} \\
& previously unselected alters are selected & \\
 & completely at random for subsequent sampling. & \\
 & $n_{cup}$ are selected whenever available.& \\
 \hline
Sample Size &  Sampling stops when $n$ nodes have been sampled. & {\bf 500}\\ [-0.3in]
\end{tabular}  \label{tab:sampparams} 
\end{center}
\end{table}

For each simulation case, we simulate 1000 networks with one RDS sample from each, and we compare five estimators, as summarized in Table \ref{tab:estimators}.

\begin{table}[h] \caption{Five estimators compared in the simulation study.}
\begin{center}
\begin{tabular}{l||c|c}
Abbreviation & Source & Estimator \\
\hline
\hline
Mean & Naive sample mean of $\vz_i$ & $\hatmu$ \\
SH &  \cite{salgheck04} & $\mush$ \\
VH & \cite{volzheck08} & $\mvh$ \\
SS & \cite{gilejasa11} & $\mh$ \\
MA & Current paper & $\mhs$ \\ [-0.3in]
\end{tabular}\label{tab:estimators}
\end{center} 
\end{table}

\subsection{Primary Results}

\begin{figure}[h]
\begin{center}
    \includegraphics[width=4in]{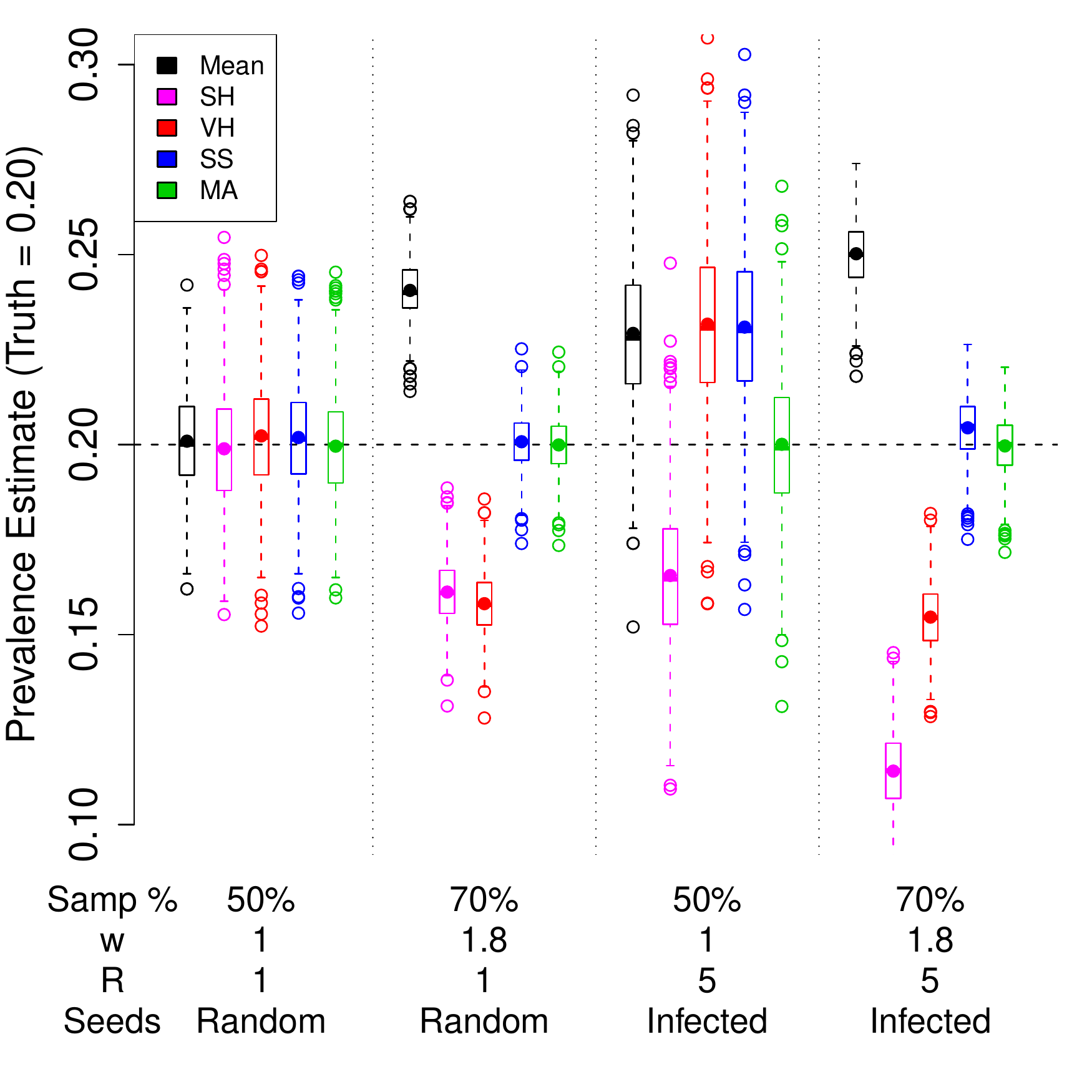}
\end{center} \vskip -0.4in\caption{Comparison of performance of five RDS estimators under four conditions.  $\mh$ and $\mhs$ assume correct population size $N$.  Results from 1000 simulations.
} \label{fig:misc1}
\end{figure}

We begin by studying the performance of the proposed estimator in settings where previous estimators have been found to perform well.  In the first part of Figure \ref{fig:misc1}, there is a relatively small sample fraction (50\%, $N=1000$), no homophily on infection status ($R=1$), the ratio of mean degrees by infection ($w$) is 1, and seeds are chosen from the full population, so there is no bias induced by seed selection.  In this case, none of the estimators considered exhibit bias, and the naive sample mean exhibits the lowest variance, although the variability is similar across estimators.

The second part of Figure \ref{fig:misc1} illustrates the case $\mh$ is designed to address.  In this case, the sample fraction is large (about 70\%, $N=715$), and infected nodes have mean degree 80\% higher than that of uninfected ($w=1.8$).  In this case there is still no homophily ($R=1$), and no seed bias.  Here, the higher-degree infected nodes are over-represented in the sample, resulting in positive bias in the sample mean.  Because of assumed linear mapping from degree to sampling probability, $\mush$ and $\mvh$ over-correct for this feature, resulting in negative bias. $\mh$ and $\mhs$ appropriately adjust for the over-sampling of infected nodes, resulting in unbiased estimators without increased variance.

The third section of Figure \ref{fig:misc1} considers the case the new estimator, $\mhs,$ is designed to address.  There is homophily ($R=5$), and all seeds are selected from among the infected nodes.  This case treats a smaller sample fraction ($N=1000$) and activity ratio 1 ($w=1$).  Here, all of the earlier estimators exhibit bias due to the selection of seeds (note the direction of bias is different for $\mush$), while the proposed estimator appropriately corrects for the selection of seeds.

The final case considers the joint effects of large sample fraction ($N=715$), non-activity ratio ($w=1.8$), homophily ($R=5$), and biased selection of seeds (all infected).  Here, the sample mean over-represents the higher-degree and initially sampled infected nodes.  $\mvh$ exhibits a strong negative bias, similar to that in the second case.  The two effects jointly cause tremendous bias in $\mush$.  $\mh$ is affected by seed bias, although not by the sample fraction.  Here, again, the proposed estimator correctly adjusts for all of these effects.  Although in this example the effects of sample fraction/activity ratio are of larger magnitude than those of seed bias/homophily, in practice the relative magnitudes of these will vary across data sets.
\begin{figure}[h]
\begin{center}
\subfigure[]
{
    \label{fig:homoph}
    \includegraphics[width=6.5cm]{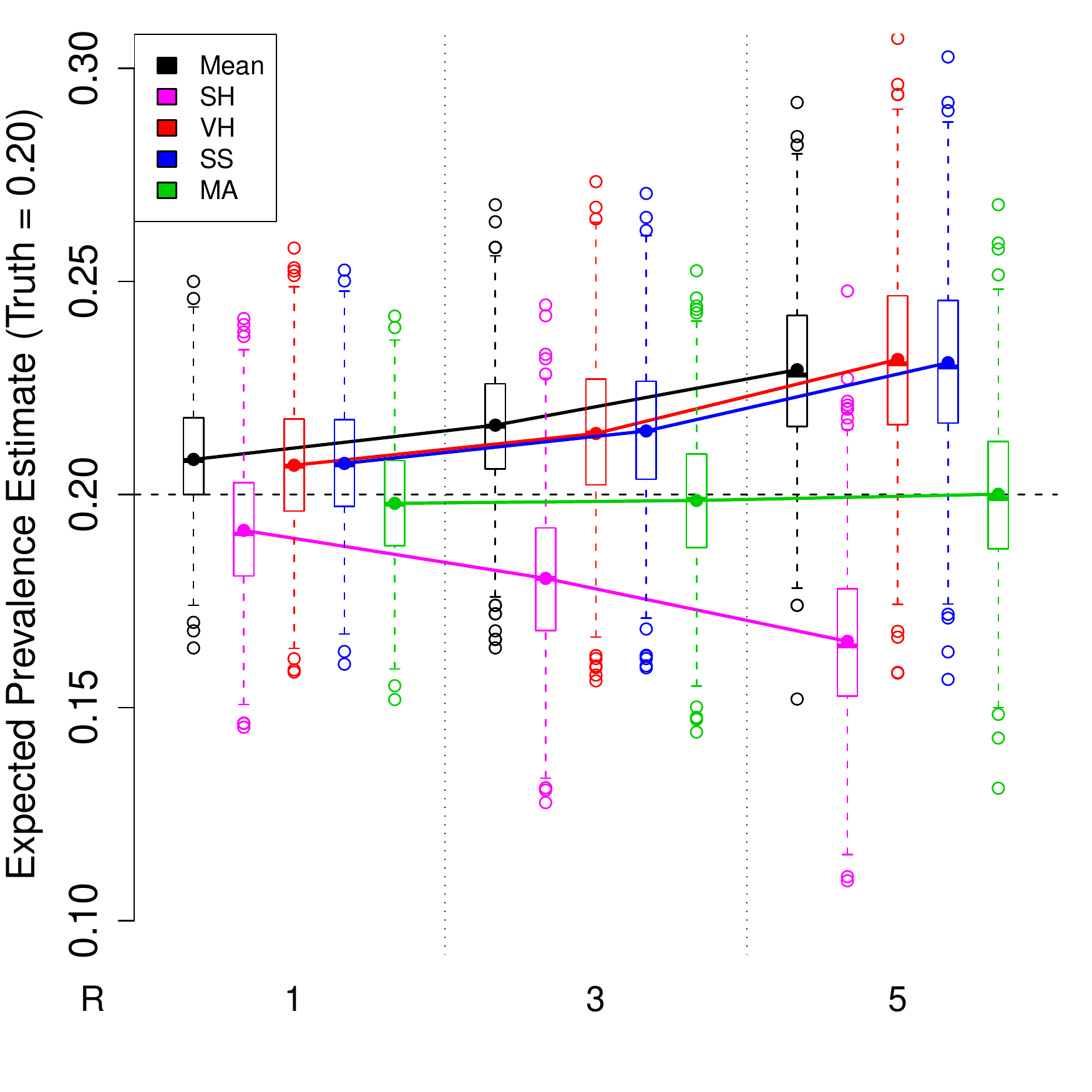}
} \hspace{-.5cm}
\subfigure[]
{
    \label{fig:waves}
    \includegraphics[width=6.5cm]{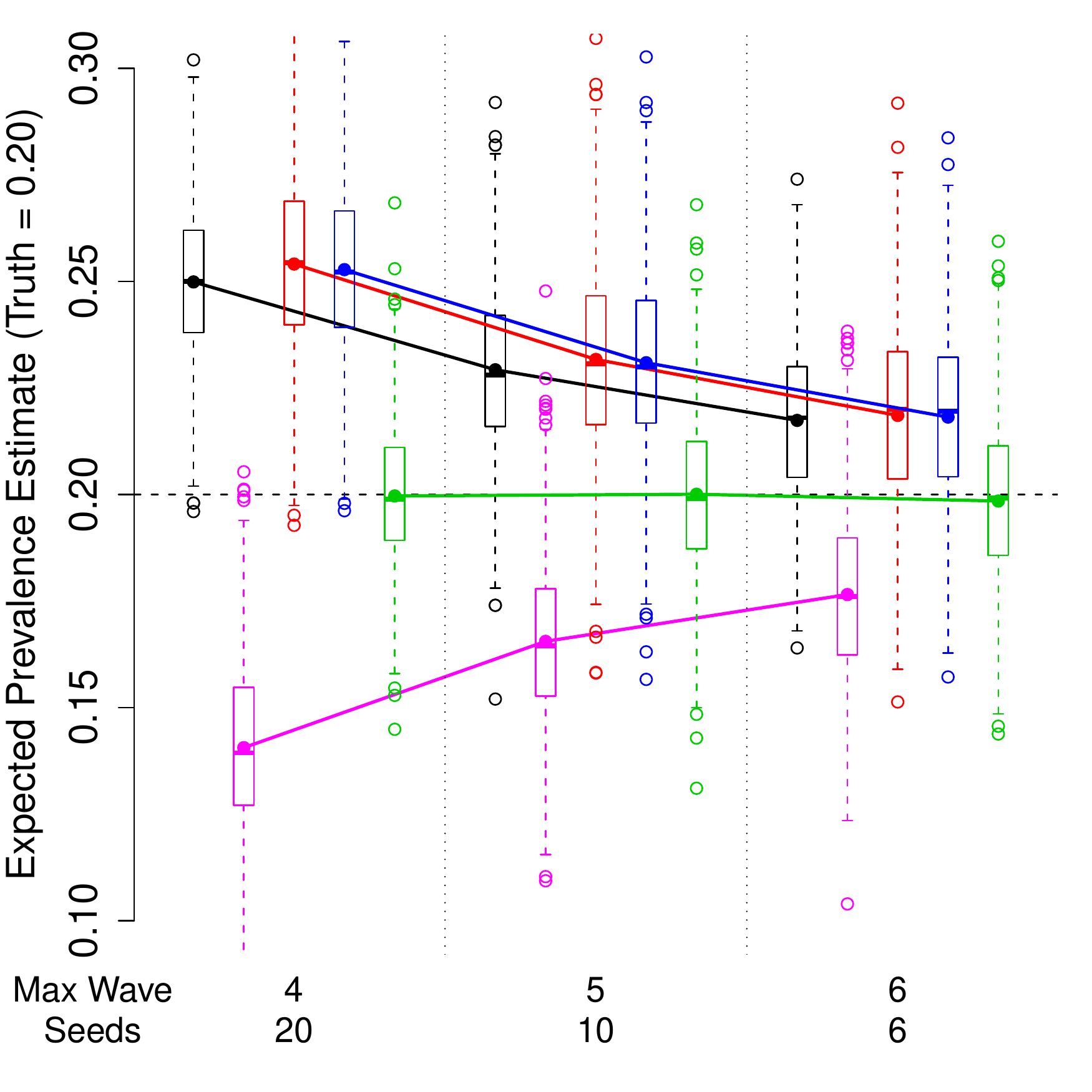}
}\end{center} \vskip -0.2in\caption{Comparison of the performance of five RDS estimators with biased seed selection and various levels of homophily and numbers of sampling waves.  All treat $N=1000, w=1$, and all seeds are selected from among infected nodes only. The first subfigure illustrates the exacerbating effect of homophily on seed bias.  The second illustrates the ameliorating effect of increased sampling waves on seed bias.
}
\end{figure}

Figures \ref{fig:homoph} and \ref{fig:waves} illustrate additional features important to the impact of seed selection on the bias of the sample mean and earlier estimators.  In particular, Figure \ref{fig:homoph} illustrates that the bias is exacerbated by increased homophily in the underlying population, and Figure \ref{fig:waves} illustrates that bias is attenuated by having more sampling waves (attained by selecting fewer seeds for fixed sample size and branching).  \knote{insert caveat in conclusions that this doesn't mean fewer seeds is better.}  In each of these cases, the proposed estimator has negligible bias.  Note that for very high levels of homophily ($R=13$), the proposed estimator was found to exhibit positive bias, but of much smaller magnitude than that of the other estimators.

\subsection{Sensitivity to the Population Size Estimate}
In practice, the size of the hidden population, $N$, may not be known.  We therefore conduct a sensitivity analysis illustrating the performance of the proposed estimator in the case of an inaccurate estimate $\hat N$ of $N$.  We consider cases of $N-\frac{1}{2}(N-n) = \hat{N}_s < N$ and $N+\frac{1}{2}(N-n) = \hat{N}_l>N$.
For each treatment of $\hat{N}$, we treat the four cases in Figure \ref{fig:misc1}.
$\hatmu$ and $\mvh$ correspond to the extreme cases of $\hat N = n$ and $\hat N = \infty,$ respectively, so are plotted for reference alongside 
$\mh$ and $\mhs$ for each level of $\hat{N}$.

Figures \ref{fig:sens1} and \ref{fig:sens2} illustrate that in the case of unity activity ratio and a smaller sample fraction, the assumed population size has little impact on the resulting estimators (note that in the case with $\hat N = 715$ in Figure \ref{fig:sens2}, seed bias leads to the perception of a non-unity activity ratio which, along with a smaller assumed population size results in the perception of a larger sample fraction).

Figures \ref{fig:sens3} and \ref{fig:sens4} illustrate that in the case of large sample fraction and activity ratio ($w\neq1$), the assumed sample fraction does impact the estimators $\mh$ and $\mhs$.  When there is no seed bias (Figure \ref{fig:sens3}), these estimators perform nearly identically.  \cite{gilejasa11} argues that in this case, $\mh$ interpolates between the sample mean and $\mvh$, and that trend seems to hold for $\mhs$ as well.   In the case of seed bias (Figure \ref{fig:sens4}), however, $\mh$ and $\mhs$ differ, in that $\mhs$ corrects for the bias induced by the seed selection.  

Finally, it is worth noting that for smaller sample fractions, such as in Figure \ref{fig:sens2}, the bias induced by seed selection may be of far greater magnitude than the bias induced in $\mvh$ by finite population effects.  For this reason, for smaller sample fractions, $\mhs$ may be able to correct for the more important form of bias, without being greatly affected by uncertainty in the population size.

\newcommand{\mouse}{5.25cm}

\begin{figure}[h]
\begin{center}
\subfigure[None]
{
    \label{fig:sens1}
    \includegraphics[width=\mouse]{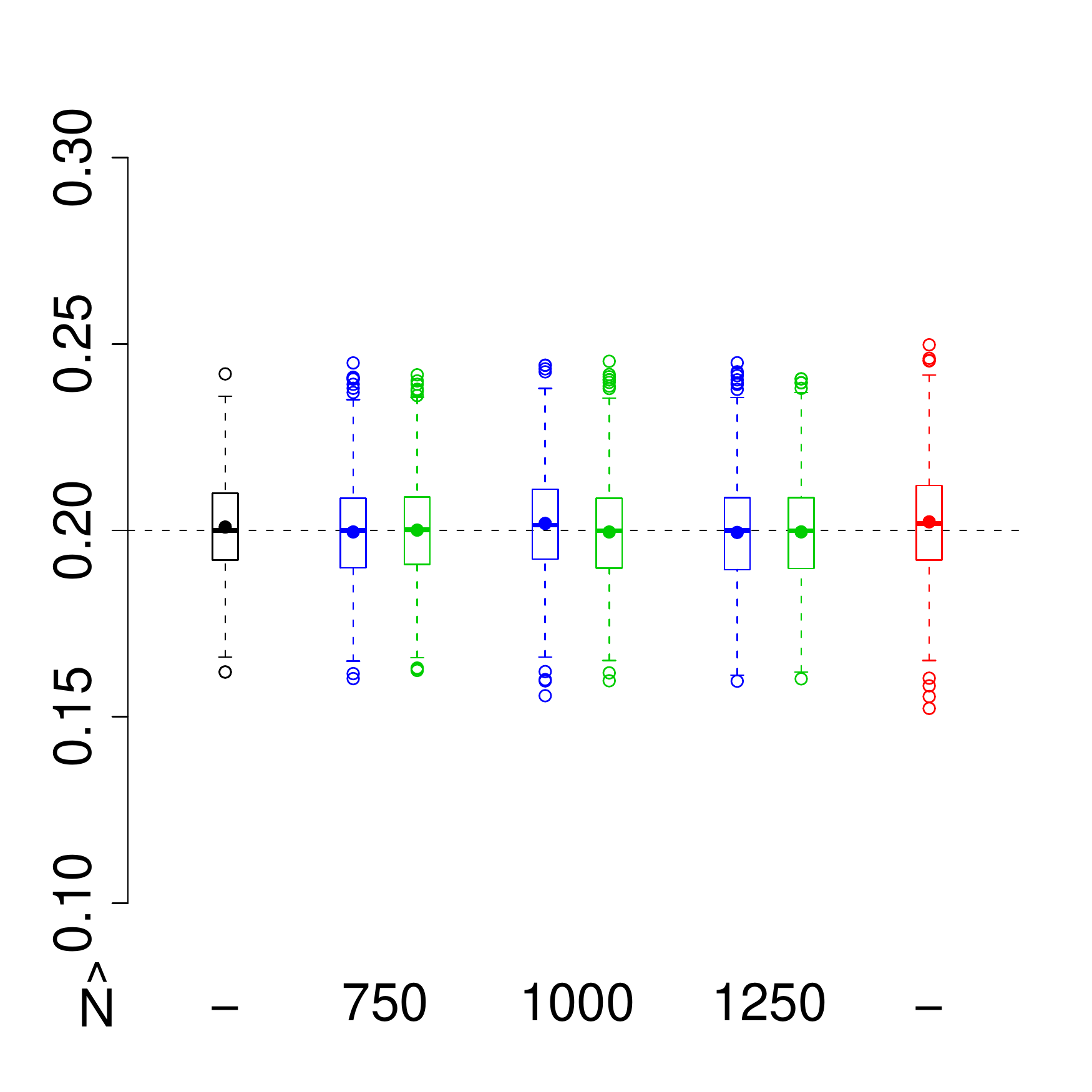}
} 
\subfigure[Sample Fraction, Activity Ratio]
{
    \label{fig:sens3}
    \includegraphics[width=\mouse]{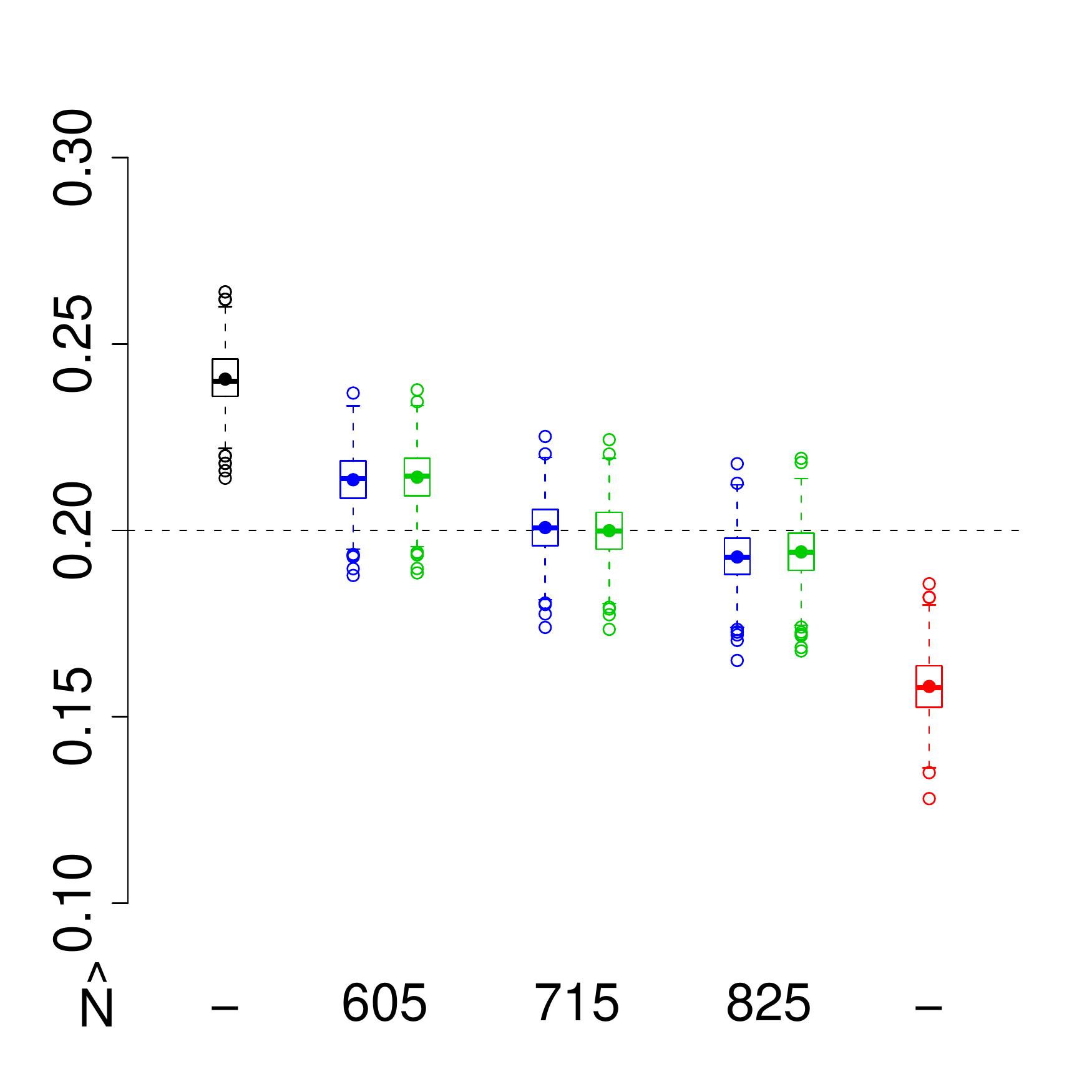}
} %
\end{center}
\end{figure}

\begin{figure}[h]
\begin{center}
\subfigure[Homophily, Seed Bias]
{
    \label{fig:sens2}
    \includegraphics[width=\mouse]{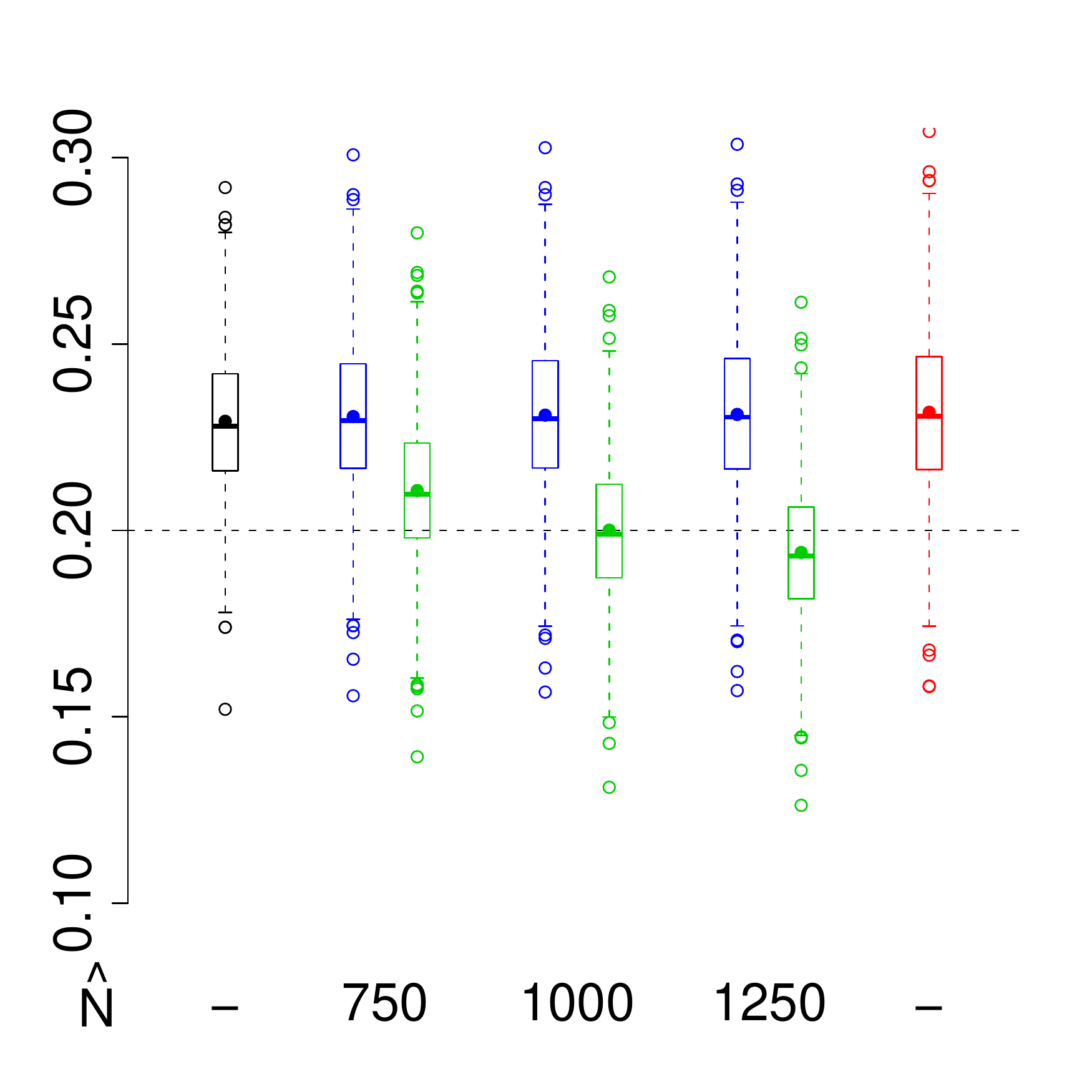}
}
\hspace{.2in}
\subfigure[All]
{
    \label{fig:sens4}
    \includegraphics[width=\mouse]{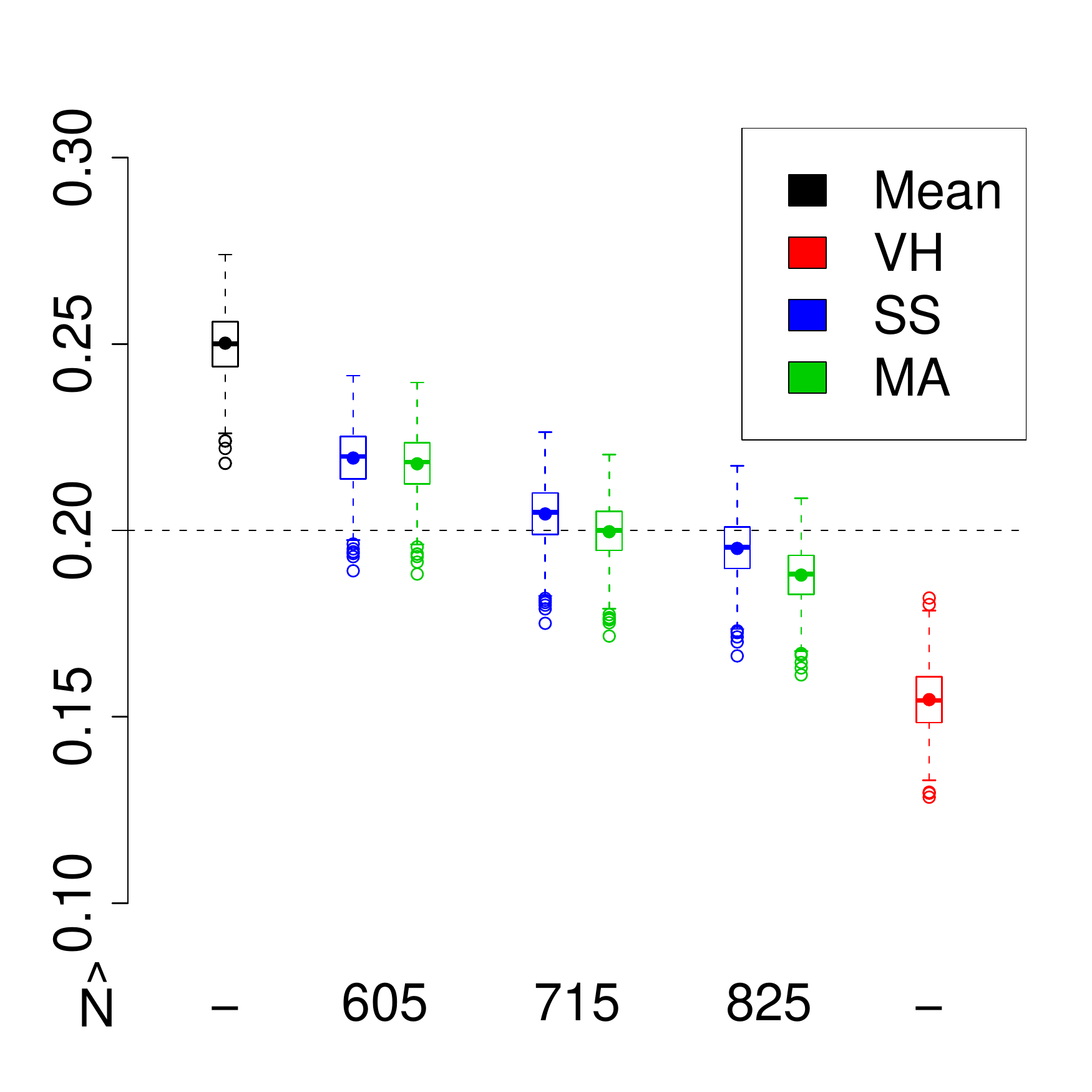}
}\end{center} \vskip -0.3in\ \caption{Sensitivity of $\mh$ and $\mhs$ to inaccurate population size under four network and sampling conditions.  \cite{gilejasa11} argues that $\mh$ approaches the sample mean for small assumed population sizes, and approaches $\mvh$ for large assumed population sizes.
}
\end{figure}

\subsection{Sensitivity to the network working model}

The role of the network working model is to provide a (stochastic)
representation of the networked population.
This model is the basis of the improved representation of the RDS design leveraged by the proposed estimator.
The complexity of real-world social networks is high, so that simple network models will typically only
capture a subset of this complexity. 

The ERGM in (\ref{ergm}) is designed to represent two levels of network
structure that are important to model RDS. The first is the nodal level representation of the individual
heterogeneity in the propensity to have social ties. This is via the nodal degrees which are also a measure of the
centrality of the individuals in the network \citep{freeman:sn:1979}. The second is at the dyadic level and
captures the homophily, or propensity for ties to be between individuals with the same infection status (beyond that implied by
the infection prevalence). As infection status is the primary outcome of interest, this homophily is
the most important to capture. The model (\ref{ergm}) does not capture third level triadic effects, those based on the structure of triads of relations between individuals.
While these are tertiary to the monadic and dyadic effects they can influence the RDS. Unfortunately RDS
results in branching tree patterns of observations that limit the empirical information on these triadic
effects. Hence the model (\ref{ergm}) presumes that the triadic effects are those that would be produced by
the modeled monadic and dyadic components.  

The purpose of this section is to assess the sensitivity of the estimator to this misspecification of the
triadic effects. 
Explicitly, we will consider networked populations with higher levels of transitivity than
specified in the network working model and compare the performance of the estimators.
Transitivity is represented by the 
edgewise shared partner (alter) statistics, denoted ${\rm EP}_{0}(\vy), \ldots, {\rm EP}_{N-2}(\vy)$,
where ${\rm EP}_{k}(\vy)$ is defined as the number of unordered pairs ${i, j}$ 
such that $\vy_{ij} = 1$ and $i$ and $j$ have exactly $k$ common alters.
It is a measure of the shared friendliness of friends. 
The geometrically weighted edgewise shared partner (GWESP)
statistic, conditional on the $\theta$ parameter, is
\[
\textrm{gwesp}_{\theta}(\vy) \equiv e^\theta\sum_{i=1}^{N-2}\left[1-(1-e^{-\theta})^i \right] {\rm EP}_i(\vy)~~~~~~~~~~~~~\theta \ge 0.
\]
The GWESP is an aggregate measure of local clustering or the overall
{\ql}inwardness\qr of ties.  The parameter $\theta$ controls just how {\ql}local\qr the
clustering needs to be.
If $\theta=0$ an edge with one shared partner counts the same as an edge with
two or more shared partners. 
If $\theta>0$ an edge with one shared partner counts {\sl less than} an edge with two or more shared partners. 
So large values of $\theta$ mean that very tight clustering is highly weighted and loose clustering is emphasized less.
These terms have been developed for ERGM by \citet{sprh06} and
\citet{hunhan06}.

\begin{figure}[h]
\begin{center}
\subfigure[Transitivity via GWESP with $\theta=0$.]
{
    \label{fig:gwesp0}
    \includegraphics[width=6.5cm]{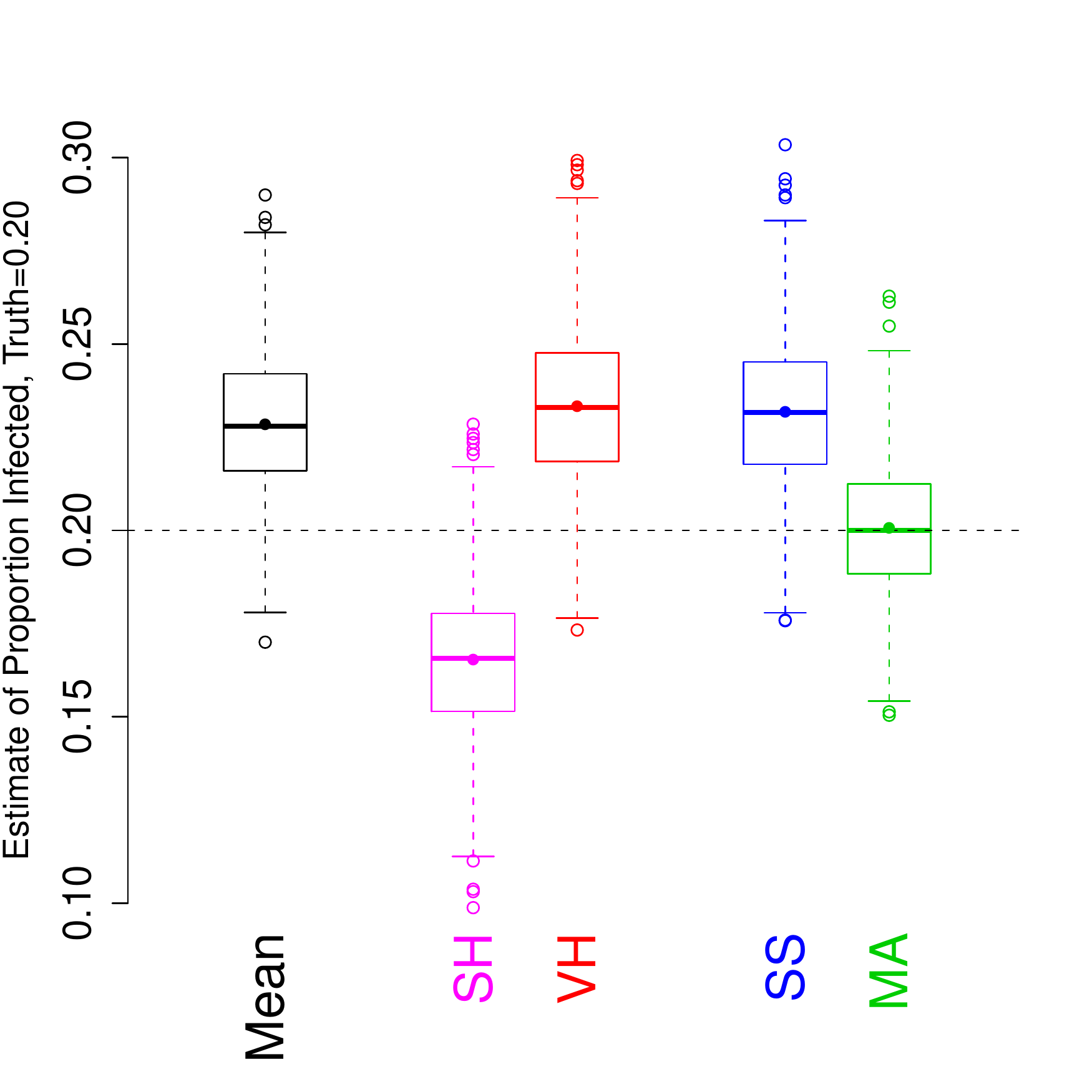}
} \hspace{-.5cm}
\subfigure[Transitivity via GWESP with $\theta=1$.]
{
    \label{fig:gwesp1}
    \includegraphics[width=6.5cm]{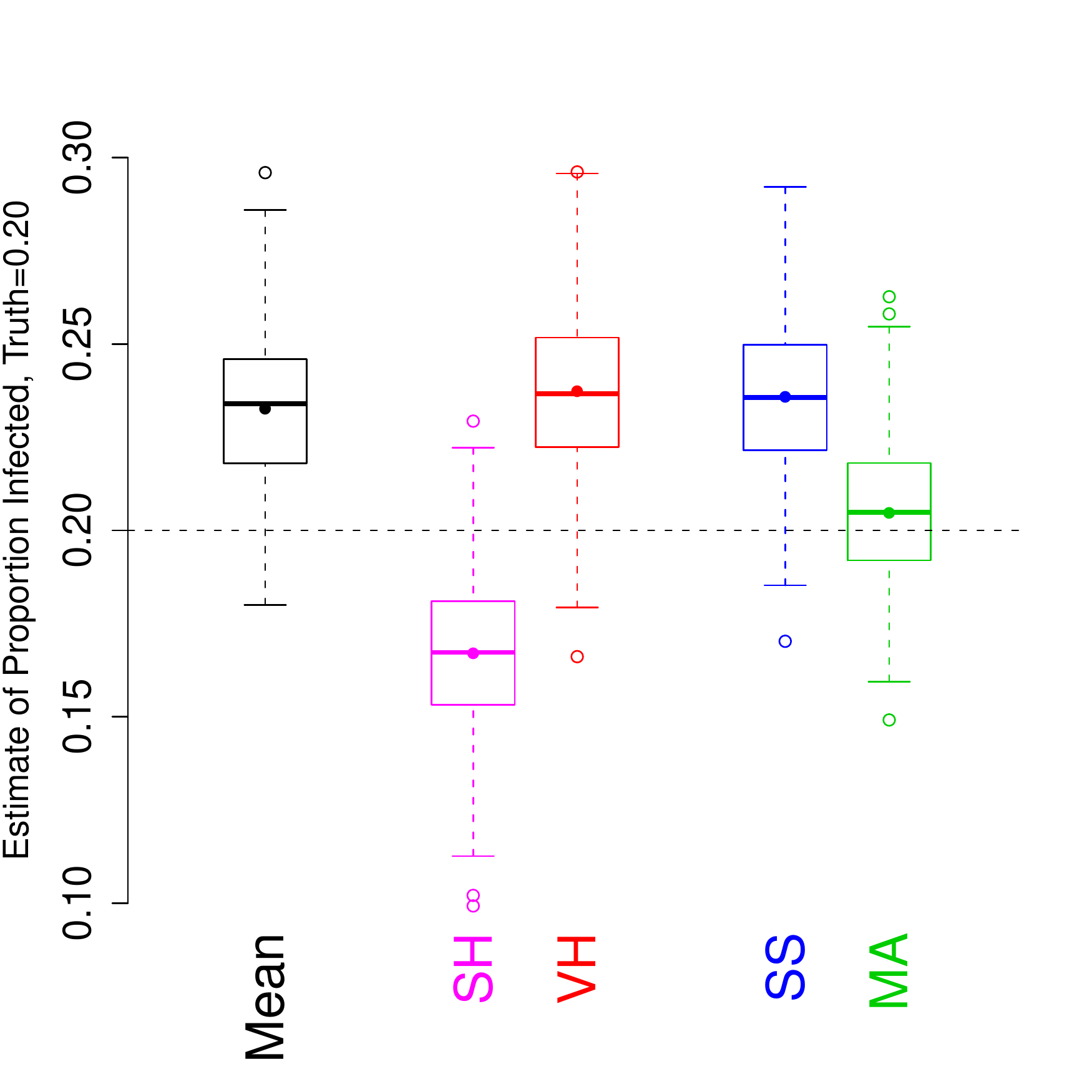}
}\end{center} \vskip -0.2in\caption{Comparison of performance of five RDS estimators when the network working model misspecifies the transitivity. The populations in the left panel have ten times the GWESP with $\theta=0$ and the right panel with $\theta=1$. Results from 1000 simulations.
}\label{fig:gwesp}
\end{figure}
Most real-world networked populations over which RDS will be applied may be expected to have higher levels of transitivity than that produced by monadic and dyadic
effects. Here we will consider two ways to produce networks with higher propensities for {\ql}friends of friends
to be friends\qrns. 
To investigate the relative performance of the estimator in populations with higher
transitivity, we generate networks with exactly the same monadic and dyadic statistics
as those considered in Section 4.2 but with higher transitivity as measured
by the GWESP statistic.
We do this by adding a $\textrm{gwesp}_{\theta}(\vy)$ term to the model (\ref{eqn:simmodel}) and inflating the mean value
parameter of the $\textrm{gwesp}_{\theta}(\vy)$ while holding the mean value parameters of the other terms, as well as the degrees of each node, fixed.
We then simulate networks from the resulting model, and apply the sampling and estimation procedures.

To test the correction for seed bias under model misspecification, we consider the populations in the third section of Figure \ref{fig:misc1}. These have a smaller sample fraction ($N=1000$), high homophily ($R=5$), no differential activity ($w=1$) and biased selection of seeds (all infected).
We take the same 1000 populations and re-generate them 
with the exactly the same degree sequences and homophily but with increased transitivity (as measured by
the $\textrm{gwesp}_{\theta}(\vy)$).

Figure \ref{fig:gwesp} compares the same estimators as in Figure \ref{fig:misc1}. The left panel
consider populations with $\textrm{gwesp}_{0}(\vy)$ ten times that in the original. 
A value of $\theta=0$ means that the statistic measures the number of
pairs of people that are connected both by a direct edge {\sl and} by a 
two-path through another person (that is, the number of edges minus the number of
edges connected by no two-paths). 
As can be seen, the performance
of $\mhs$ is little effected by the increased transitivity. 
The right panel compares populations with 
ten times $\textrm{gwesp}_{1}(y).$
A value of $\theta=1$ means that the statistic weighs up the
connectedness of edges with more weight on the terms with more
shared partners.  
In this case $\mhs$ has modest positive bias (0.46\%) and similar variance
compared to the estimators on the original populations.

\section{Application to HIV prevalence in Hidden Populations}
\label{sec:application}

We apply our estimator to data collected in 2007 among injecting drug users (IDU) in Mykolaiv, Ukraine.  The HIV epidemic in the Ukraine is one of the most severe in Europe, and still growing.  As of 2009, the adult HIV prevalence was estimated at 0.86\% \citep{ukraine2009}.  Ukraine's epidemic is most severe among injecting drug users and their sexual partners, who account for the majority of new infections \citep{USAID2010}.  The data we consider here were collected as part of a series of studies of IDU across major Ukrainian cities in 2007 \citep{Kruglov08}.  We focus on the data collected in Mykolaiv because, by chance and because of the contacts available to the researchers, all seeds in this sample were HIV positive.

This study began with 6 seeds and continued until wave 10, with 31 samples from wave 10, and a total of 260 samples.     
The average wave number was 6.1. The homophily based on HIV status for the population is estimated to be $R=2.47$ and the differential activity is estimated to be 0.72.

Although the size of the population was not known precisely, an estimated range of population sizes is available through scale-up and multiplier methods \citep{Kruglov08, UNAIDS2003}.  We chose a population size, $N=4000$, near the low end of this range.  The variability of population size estimates is quite large, with a point estimate closer to 8000 in 2008 \citep{berleva2009}).  We used sensitivity analysis to verify that population size 4000 is sufficiently large that our estimates are insensitive to increased population size.  

\begin{figure}[h]
\begin{center}
    \includegraphics[width=3.75in]{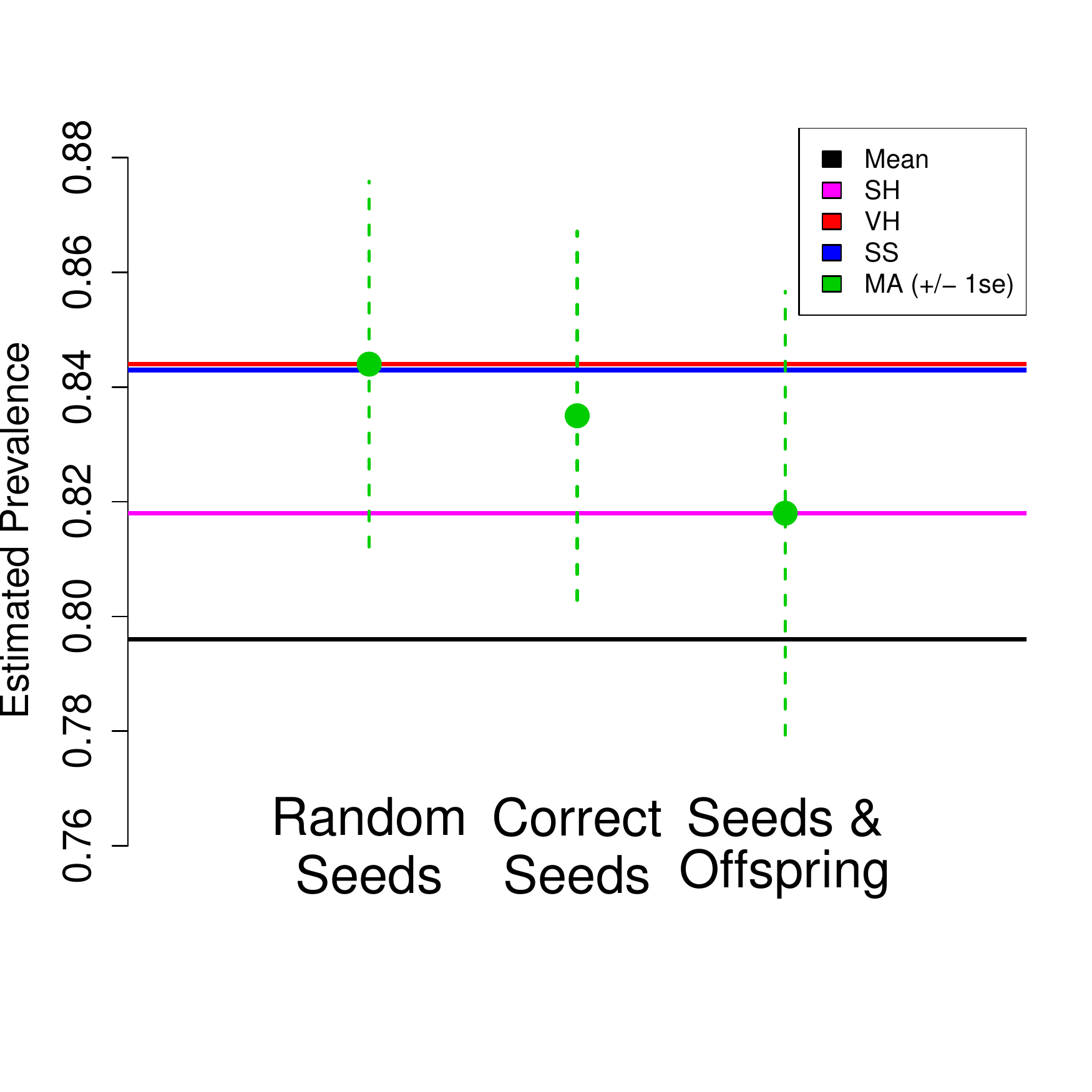}
\end{center} \vskip -0.7in\caption{HIV prevalence estimates for injecting drug users in Mykolaiv, Ukraine, 2007.} 
 \label{fig:myk}
\end{figure}
We compare estimates for this application based on the current standard estimators ($\mush$, $\mvh$, $\mh$) and three variants of $\mhs$, summarized in Figure \ref{fig:myk}.  First, we consider a version of $\mhs$ which does not correct for seed bias.  In this case, we select the seeds of the simulated samples with probability proportional to degree and without regard to infection status.  As illustrated in Figure \ref{fig:myk}, this results in an estimate very close to that given by $\mh$ and $\mvh$ ($\mvh = 0.844$, $\mh = 0.843$, $\mhs = 0.845$).  In the second condition, we then apply the correction for seed bias, by matching the simulated seeds to the infection and degree characteristics of the observed seeds.  This results in the second estimate in Figure \ref{fig:myk}, $\mhs = 0.837$.  This adjustment is in the direction we would expect, decreasing the prevalence estimate, corresponding to down-weighting the group over-represented in the seeds.  The modest magnitude of this adjustment can be attributed to the weak homophily in this network, relative to the number of sample waves, as well as the high prevalence, leading to a smaller difference between prevalence in the population and prevalence among the seeds than in our simulation study. 

\begin{figure}[h]
\begin{center}
    \includegraphics[height=3in]{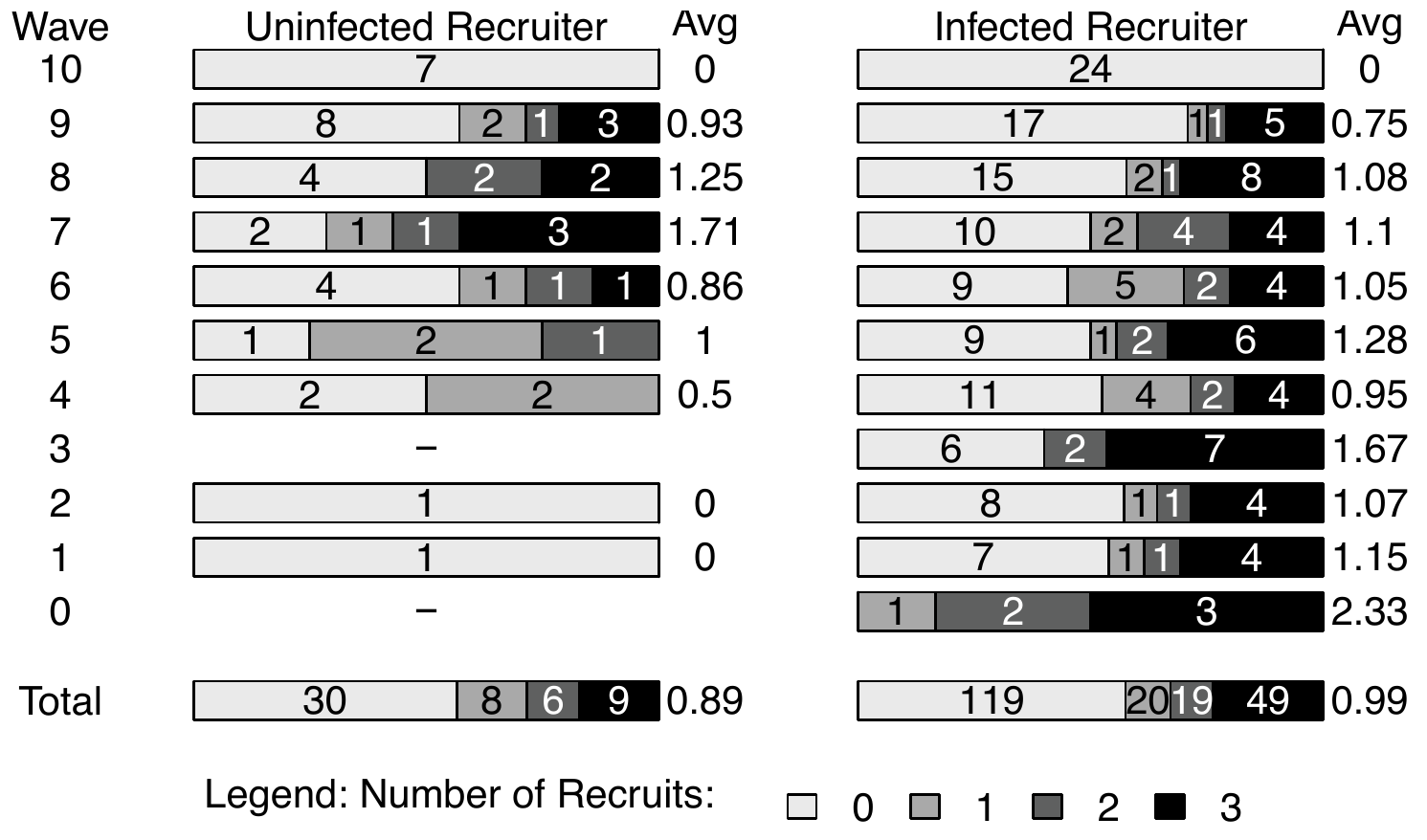}
\end{center} \vskip -0.2in\caption{Distribution of number of successful recruitments by wave and infection status of recruiter.  Uninfected recruiters were rare and unsuccessful in early waves, contributing to under-representation of uninfected participants in the sample.  There were no uninfected participants in waves 0 (seeds) or 3.
} \label{fig:mykrecruitbars}
\end{figure}

The third condition we considered highlights the flexibility and possibilities for extensions of $\mhs$.  We note that in these data, infection groups differed in their recruitment behavior.  Some differences in recruitment behavior have been referred to as differential {\it recruitment effectiveness} in \cite{heck07}, as well as in \cite{tomasgile} and \cite{aditya}.   This is a pattern in which one group systematically recruits more effectively than another group.  In this case, however, the pattern was more complex.  On average, infected and uninfected participants did not vary greatly in their recruitment effectiveness. However, uninfected participants {\it in the early waves} of the study recruited disproportionately few additional participants, as illustrated in Figure \ref{fig:mykrecruitbars}.  Because of the branching nature of the sampling, this resulted in a dramatic under-representation of uninfected IDU in the survey.  To correct for this, however, we needed to estimate and replicate offspring distributions varying by both infection status and survey wave.

We therefore applied a version of $\mhs$, modified to reflect the empirical offspring distribution by wave and infection status.  In most cases, this required simply assigning an offspring distribution equal to the empirical offspring distribution by wave and infection status.  For uninfected recruiters in wave 3, we used averaged empirical values from waves 2 and 4.  For waves 10 and beyond, we replicated the empirical results from wave 9.  Whenever a simulated recruiter did not have enough eligible alters to allow for the number of recruits selected from the appropriate distribution, we assigned any unfulfilled recruitments to the next active recruiters of the same infection status with fewer than 3 assigned recruits.  
This is straight-forward to apply in our model-assisted setting and illustrates how this approach allows the
approximation to the sampling design to be improved using available information.

The results of this analysis are illustrated in the third bar of Figure \ref{fig:myk}.  The resulting estimate, $0.817$, was substantially lower than the earlier estimates, suggesting the offspring distribution had a substantial impact on the resulting estimates.

\section{Discussion}

In this article we introduce a new approach to estimation based
on RDS data that uses a working model for the underlying
networked population to more accurately estimate the inclusion
probabilities necessary for design-based inference.

We demonstrate that this approach allows us to correct for differential sampling probabilities based on nodal degrees, as in earlier RDS estimators \citep{salgheck04, heck07, volzheck08}, as well as for finite population biases, as addressed by another earlier approach \citep{gilejasa11}.  In addition, our proposed estimator is able to adjust for the convenience sample of seeds, a feature not accounted for in any previous approaches.  

We apply this approach to obtain improved estimation of HIV prevalence in an IDU population in the Ukraine. 
We improve the approximation to the actual RDS process resulting in improved estimates,
and compute associated measures of uncertainty.
We also show the flexibility of the working model approach. It allows for additional information available in a particular
application to be incorporated via the ERGM framework, and leverages recent advances in that area \citep{statnetjss,sprh06}.
A significant weakness of our approach is the requirement that the population size is known.  Our simulation study illustrates that the proposed estimator is indeed sensitive to estimates of population size, but as long as the population size is not greatly under-estimated, we do not expect it to perform worse than the earlier estimator $\mvh$,  and in cases of bias induced by the sample of seeds, $\mhs$ may perform substantially better than any existing estimator, even for highly inaccurate assumptions regarding the population size.

We therefore propose three practical regimes for the use of $\mhs$.  First, if the population size is known, the estimator may be applied directly.  If the population size is unknown, but a range of estimates is available, the estimator may be applied across the range.  If the results vary greatly, the uncertainty of the resulting estimator should be adjusted accordingly.  Such may be the case, for example, in populations of men who have sex with men, who are assumed to constitute 1\% - 3\% of many populations.  Finally, if no information on population size is available, $\mh$ may be compared with $\mvh$ to determine whether there are important finite population effects in this sample.  If not, $\mhs$ may then be applied to correct for any effects of seed bias.  In the case where finite population effects are found, $\mhs$ may be applied to diagnose the extent of seed bias at each of several estimates of population size.  Note that in such cases, the earlier estimators also make an assumption about population size (i.e. that it is {\it sufficiently large}), and so do not avoid the problem of unknown population size.

Another important assumption is the form of the social network working model.  Our estimator relies on a simple model, not because we believe it to be strictly accurate, but because we expect it to capture the network features most important to the sampling process, and because it is feasible to estimate from the available data.
To assess the sensitivity of the estimator to the form of the working model we considered versions of
populations with greatly increased transitivity, a feature not captured by the working model. The results
indicate only modest impact of high transitivity on the estimator or it uncertainty. While this may not be
universally true, it does indicate the ability of the working models to capture nodal and dyadic effects goes a
long way to improve the representation of the RDS process.

Several extensions of this approach are possible.  First, if data on the characteristics of all alters are not available, we may wish to estimate the sum of cross-group ties ($\vm$)  based on referral patterns.  Such an estimate is used in the application to HIV prevalence estimation (Section \ref{sec:application}).

Our approach can also be extended to include additional measurable features of the network working model or sampling process, such as homophily on neighborhood of residence or bias in the passing of coupons.  We illustrate one such  extension in Section \ref{sec:application}, in which we observe an aberrant pattern of recruitment by infection status, and adapt the estimator to condition on this pattern.  Note that the resulting estimate is very close to that given by $\mush$.  This is consistent with results in \cite{tomasgile} indicating that $\mush$ is not as susceptible to bias induced by differential recruitment effectiveness as $\mvh$ or $\mh$.

Further extensions of this approach will make it possible to consider the joint estimation of population size and prevalence, or correlations between multiple nodal variables.  We explore these features in ongoing work.

We intend to make code available for these procedures in the R package {\tt RDS} on CRAN \citep{RDS, R}.

\begin{description}
    \item[Inferential and Computation:]This supplement presents specifics of the estimation algorithms and
our approach to standard error estimation (RDSMAsupplement.pdf)
\end{description}

\blind{{\relax}{
\section*{Acknowledgements}
The project described was supported by grant number 1R21HD063000 from NICHD and
grant number MMS-0851555 from NSF, and grant number N00014-08-1-1015 from ONR. Its contents are solely the responsibility
of the authors and do not necessarily represent the official views of the
Demographic \& Behavioral Sciences (DBS) Branch,
the National Science Foundation, or the Office of Navel Research.
The authors would like to thank
the members of the Hard-to-Reach Population Research Group (\url{hpmrg.org}), especially Lisa G. Johnston and Cori M. Mar, for their helpful input, and also to thank Tetiana Saliuk of AIDS Alliance, Ukraine for the use of her data in the application.
}}

\advance\baselineskip by -6.8pt
\advance\parskip by -7pt
\addcontentsline{toc}{section}{References}
\bibliographystyle{JASA_manu}
\bibliography{networksjasa}

\end{document}


\maketitle

\newcommand{\N}{\mathbb{N}}
\newcommand{\s}{\mathscr{S}}

\newcommand{\q}{Q}
\newcommand{\qdn}{Q_{\N,n}}
\newcommand{\qys}{\q^*_{k,z}(y,\s)}
\newcommand{\qysk}{\q^*_{k,z}(y^i,\s^i)}
\newcommand{\mI}{\hat{\mu}_I}
\newcommand{\hatmu}{\hat{\mu}}
\newcommand{\mush}{\hat{\mu}_{SH}}

\newcommand{\p}{{\bf p}}
\newcommand{\vp}{{\bf p}}
\newcommand{\D}{{\bf D}}
\newcommand{\X}{{\bf X}}
\newcommand{\Z}{{\bf Z}}
\newcommand{\vZ}{{\bf Z}}
\newcommand{\Y}{{\mathscr{Y}}}
\newcommand{\vd}{{\bf d}}
\newcommand{\z}{{\bf z}}
\newcommand{\vz}{{\bf z}}
\newcommand{\vs}{{\bf s}}
\newcommand{\vS}{{\bf S}}
\newcommand{\vV}{{\bf V}}
\newcommand{\vv}{{\bf v}}

\newcommand{\zs}{{\bf z_s}}
\newcommand{\ds}{{\bf d_s}}  
\newcommand{\pc}{{\bf p_c}}

\newcommand{\mhs}{\hat{\mu}_{MA}}
\newcommand{\piihat}{\hat{\pi}_{i}}

\section{Estimation Procedure}\label{estpro}

We propose the following algorithm to compute the new estimator $\mhs$ of $\mu$. 
\begin{enumerate}
  \item Estimate the following according to their empirically observed values:
  \bi
    \item Sample size $n$
    \item Number of seeds $n^{seeds}$, and degree and infection status of seeds, given by $\N^{seeds}$ = $\{\N^{seeds}_{ij}\}$, where $\N^{seeds}_{ij}$ represents the number of seeds with degree $i$ and infection $j$, $i \in 1 \ldots N-1$, $j \in \{0,1\}$.
    \item Offspring distributions $\vp^{s}$, where $\vp^{s}_i = $ the proportion of the sample with $i$ offspring, $i = 0, 1, \ldots, $ {\rm maximum number of coupons}.
  \ei
  \item Estimate: 
  \[
   \piihat^0 = \frac{d_i}{N} \sum_{j=1}^N \frac{S_j}{d_j}, ~~~~~ i: S_i = 1.
  \]
  \item For $r = 1 \ldots h$:
  \begin{enumerate}
    \item Estimate:
   \begin{align*}
  \tilde{\N}_{kl}^r &=  {1\over{N}}\sum_{i=1}^{N} \frac{\vS_i \mathbb{I}({\vd_i=k, \vz_i=l})}{\piihat^{r-1}} \label{nhatr}\\
  \mtilde^r &=  \sum_{i=1}^{N} \frac{\vS_i \left( \vx_i (1-\vz_i) + (\vd_i - \vx_i) \vz_i   \right)}{2 \piihat^{r-1}} 
\end{align*}
  \item Compute the ERGM parameter $\eta$ in the model (2) based on ${\tilde \N}^r$ and $ {\mtilde}^r$ via the procedure in Supplemental Section 3. Denote the estimate by $\eta^r.$ 
  This step is conducted using the {\tt statnet} R package \citep{statnet}.
  \item Simulate $M_1$ networks according to the distribution given by $\hat{\eta}^r, {\tilde \N}^r,$ and ${\mtilde}^r$, also using the {\tt statnet} R package.
  \item Simulate $M_2$ RDS samples from each of the $M_1$ networks in the previous step, according to sampling parameter $\s = \{n, \N^{seeds}\},
  \vp^{s}$.  Let $U_{kl}^r$ represent the number of times a node of degree $k$ and infection $l$ is sampled, over all $M = M_1 \times M_2$ samples.
  \item Estimate $\piihat^r ~ \forall~ i: \vS_i =1$ in a manner similar to \cite{fattorini06} and \cite{gilejasa11}:
  \[
    \piihat^r = \frac{U_{\vd_i \vz_i}^r  + 1}{M \cdot \N_{d_i z_i}^r +1}
  \]
  \end{enumerate}
\item Let $\piihat = \piihat^h$
\item Estimate
\begin{equation}
\mhs = \frac{\sum_{i=1}^N \frac{\vS_i \vz_i}{\piihat}}
                     {\sum_{i=1}^N \frac{\vS_i}{\piihat}}\label{mhs}
\end{equation}
\end{enumerate}
The simulations in this paper are based on $r=3$ iterations, each including $M_1=25$ network samples and $M_2=20$ RDS samples from each network.
In general, we recommend at least $M_1=25,$ $M_2=20$ and $r = 3$.
Estimation time scales with sample size, population size, and M. In
our simulations, with $N = 1000$, estimates require about 20 minutes on a personal computer.
In practice, these parameters can be adjusted for desired
precision in the solution to (7).
We will make available the code to compute this estimator in the {\tt RDS} R package on CRAN \citep{RDS, R}.

\section{Measures of Uncertainty: Bootstrap}
\label{bootstrap}

Unlike earlier RDS estimators, the estimator given in
(7) 
allows for estimators of uncertainty that account
for the estimated full relational structure of the underlying
population as well as incorporating several observable features of
the sampling process.  The former is because of the use of the
network working model for the population over which the RDS
sampling procedure operates.  The latter is because our procedure
enables the simulation of complex RDS designs.  In particular, if we
believe there is seed bias we can incorporate it into the
sampler, and if there is measurable sampling bias (as in the application) we
can incorporate that also.  This allows the procedure to
incorporate available information about the population and
sampling and greatly improves the accuracy of the representation of
the actual sampling process. The accuracy of the bootstrap depends directly on the quality of the approximation to the actual sampling process.

\newcommand{\Dmhs}{\hat{\mathscr{D}}(\mhs)}

We propose a parametric bootstrap approach to obtaining confidence intervals, according to the following procedure:
\begin{enumerate}
\item For $b=1, \ldots B$, iterate the following steps:
\begin{enumerate}
\item Simulate a network $\vY_b$ from the model given in (2) 
with parameters $\eta = \hat{\eta}^h$, 
and $\N^h$ 
where $h$ is the final iteration of the algorithm in Supplemental Section 1. 
\item Simulate one RDS sample $S_{\vY_b}$ with parameter $\s_b$ from $\vY_b$.
\item Compute an estimator $\mhs(b)$ of $\mu$ based on the sample $S_{\vY_b}$ using the algorithm in Supplemental Section 1. 
\end{enumerate}
\item Use the empirical distribution $\Dmhs$ of $\{\mhs(1), \mhs(2), \ldots \mhs(B)\}$ to estimate the distribution of $\mhs$ under the estimated model form.
\end{enumerate}

The distribution of $\Dmhs$ may then be
used to form confidence intervals for $\mu$ which account for the
full estimated relational structure as well as observable biases
of the sampling process. 
We use the standard deviation of the resulting population of $B$
bootstrap estimates as an estimate of the standard error of
$\mhs$.  We have used $B=1000$ bootstrapped samples.  In our
simulations, this procedure took about 20 minutes per sample on a
single processor.  Parallelization is straightforward and
dramatically reduces elapsed time.
A large additional speedup can be obtained by replacing step (c) with (c') in which the weight {\it classes}
$\piihat^h$ from (\ref{mhs}) are reused to weight each bootstrap sample. While these weights vary from bootstrap
sample to sample, their uncertainty is a small part of the overall uncertainty. This reduces the procedure to 
about 30 seconds per sample on a single processor. The analysis below uses (c').

We illustrate the performance of this standard error estimator by comparing five critical cases.  As with the point estimate, we illustrate both cases in which we expect the estimator to perform reasonably well, and a case in which we expect the estimator to perform poorly.
We introduce various forms of sampling biases.
The initial sample can be selected either independent of infection status
(denoted {\ql}No\qr in the bias column of Table 1) or all from within the infected subgroup
({\ql}Initial\qr bias). We  also introduce referral bias where all infected
alters are 20\% more likely to be referred than uninfected
alters ({\ql}Referral\qr bias).

Each set of simulations involved 1000 bootstrapped re-samples for each of 1000 simulated RDS samples.  The parameters of the samples, average estimated standard errors, and coverage rates of nominal $95\%$ and $90\%$ confidence intervals are given in Table 1. 

\begin{table}[h]\caption{Observed (simulation) standard errors of estimates, and average bootstrap standard error estimates, along with coverage rates of nominal 95\% and 90\% confidence intervals for procedure given in Supplemental Section 2 
for varying sample proportion, homophily $R$, and activity ratio $w$, and for
various biases in the sample selection process.
Observed standard errors are based on 1000 samples.
Bootstrap standard errors are the average bootstrap standard
error estimates over the same 1000 samples.  Nominal confidence
intervals are based on quantiles of the Gaussian distribution.}

\begin{center}
\begin{tabular}{cccc|cc|cc}
\% & homoph. & & sample & SE & SE & coverage & coverage \\
sample & R & $w$ & bias & observed & bootstrap & 95\% & 90\%\\
\hline
\hline
50\% & 1 & 1 & No        & 0.0140 & 0.0137 & 94.1\% & 88.8\% \\
70\% & 1 & 1.8 & No      & 0.0073 & 0.0075 & 94.9\% & 90.4\% \\
50\% & 5 & 1 & Initial   & 0.0188 & 0.0175 & 93.7\% & 87.9\% \\
50\% & 5 & 1.8 & Initial & 0.0079 & 0.0080 & 95.0\% & 87.3\% \\
50\% & 5 & 1 & Referral  & 0.0216 & 0.0225 & 91.7\% & 84.7\% \\
\end{tabular}
\end{center}\label{tab:boot}
\end{table}

The magnitudes of the average bootstrap standard error estimates are quite close to the observed values in the first four cases, and the coverage rates in the cases without referral bias are very close to their nominal values.  In this last case, the standard error estimator is anti-conservative because the bootstrap procedure does not replicate the referral bias in the sample. 

The last row of Table 1 
illustrates the poor performance of the estimator in the case of extreme referral bias.  In this case, the estimator $\mhs$ has positive bias ($1.74\%$), leading to moderately lower coverage rates of the nominal intervals.

\section{Inference for the ERGM conditional on the degree and infection status sequences}


The model-assisted approach is based on a {\ql}working\qr model (2)\ 
for the networked population.  The unknowns in the model are the
finite-population values $\vd$ and $\vz$ and the
super-population parameter $\eta.$ 
Finite-population estimates of $\N$ (i.e., $\vd$ and $\vz$) and $\vm$ are determined by 
design-based inference as in (5) and (6).
The estimate of $\eta$ is 
computed as the natural parameter in (2) 
corresponding to these values.
That is, the natural parameter corresponding to the
mean-value parameter ${\mtilde}$ conditional on the degree
sequence $\vd$ and infection status sequence $\vz$ induced by ${\tilde \N}$.
Explicitly, we construct the joint degree and infection status 
sequence $\vd, \vz$ from $\N,$ where the ordering of nodes is arbitrarily assigned
(w.l.o.g.). To compute $\eta$ we 
construct a network with this
joint degree and degree status sequence and cross-group contacts
$g(\vy, \vz)$ using the Reed-Molloy method and then simulated
annealing \citep{molloyreed1995,statnetjss,statnet}.

We can then compute ${\hat\eta}$ using the Geyer-Thompson MCMC approach
\citep{statnetjss,statnet}.  As this is computationally expensive
and unstable in this situation we use an approach based
on a form of pseudo-likelihood introduced below.

Consider a model similar to (2) but with network space $\mathscr{Y}$
consisting of all binary undirected networks (i.e.,  
unconditional on $\vd$ and $\vz)$.
Until recently inference for such models 
have been almost exclusively based on a local alternative to the likelihood
function referred to as the \textit{pseudo-likelihood}
\citep{Bes75, Str90}.
Consider the conditional formulation of this model: 
\begin{equation}\label{eq:autolog}
{\rm logit[}P(Y_{ij} = 1{\vert} Y_{ij}^c = \vy_{ij}^c, \eta)] =
\eta\delta(\vy_{ij}^c )\quad \quad \quad \vy \in {\mathscr{Y}}
\end{equation}
where $\delta (\vy_{ij}^c, \vz) = g(\vy_{ij}^+, \vz) - g(\vy_{ij}^-, \vz),$ the change
in $g(\vy, \vz)$ when $\vy_{ij}$ changes from 0 to 1 while the remainder of
the network remains $\vy_{ij}^c$ \citep[See][]{Str90}.
The pseudo-likelihood for the model is: 
\begin{equation}
\ell_{P}(\eta;\vy) \equiv
\eta{\sum_{ij}{\delta (\vy_{ij}^c, \vz) \vy_{ij}}}
- \sum_{ij}\log\left[1+\exp(\eta\delta (\vy_{ij}^c, \vz))\right].
\label{eq:pl}
\end{equation}
This is the standard form of pseudo-likelihood, which we refer to a dyadic
pseudo-likelihood. 

This form is algebraically identical to the likelihood for
a logistic regression model where each unique element of the
adjacency matrix, $\vy_{ij}$, is treated as an independent observation with the
corresponding row of the design matrix given by $\delta(\vy_{ij}^c, \vz)$.  Then the maximum likelihood estimate (MLE) for this
logistic regression model is identical to the maximum dyadic pseudo-likelihood
(MPLE) for the corresponding ERG model, a fact that is exploited in computation.
Therefore, algorithms to
compute the MPLE for ERGMs are typically deterministic while the algorithms
to compute their MLEs are typically stochastic.
In addition, algorithms to compute the MLE can be unstable if the model is
near degenerate \citep{handcock:tr:2003}. This can lead to computational failure.

This standard form of pseudo-likelihood is inappropriate for the model
(2) 
as it does not take into account the network space $\mathscr{Y}(\vz, \vd).$
This is because  $P(\vY_{ij} = 1{\vert} \vY_{ij}^c = \vy_{ij}^c, \eta)$ is either
$1$ or $0$ depending on if the value of $\vy_{ij}$ because the model conditions on the degree 
sequence consistent with $\vd$. 
Hence the MPLE will usually produce non-sensical results.

Instead of a dyadic pseudo-likelihood we develop a tetradic pseudo-likelihood.
We focus on ordered dyad-quads $\vy_{ijkl}=(\vy_{ij}, \vy_{kl}, \vy_{il}, \vy_{jk})$
such that $\vy_{ij}=\vy_{kl}=1, \vy_{il}=\vy_{jk}=0.$
We refer to this configuration as $\vy_{ijkl}^+$. For each such dyad-quad there exists an
alternative realization in which $\vy_{ij}=\vy_{kl}=0, \vy_{il}=\vy_{jk}=1.$ 
We refer to this configuration as $\vy_{ijkl}^-$.
Thus $\vy_{ijkl}^+$ and $\vy_{ijkl}^-$ represent 
a pair in which $\vy_{ij}$ is toggled from $1$ to $0$ in such a way as to retain
the degree and infection status sequences of the corresponding full network.


Let $\vY_{ijkl}=(\vY_{ij}, \vY_{kl}, \vY_{il}, \vY_{jk}),$ 
$\vY_{ijkl}^c = \vY \backslash \vY_{ijkl}$ and
$\vy_{ijkl}^c = \vy \backslash \vy_{ijkl}$.
Let ${\mathscr D} = \{ ijkl : \vy_{ijkl}^c \cup\vy_{ijkl}^{+} 
\in {\mathscr{Y}(\vz, \vd)} \}.$
For these dyad-quad configurations we then have:
\begin{equation}\label{eq:tautolog}
{\rm logit[}P(\vY_{ijkl} = \vy_{ijkl}^{+}{\vert} \vY_{ijkl}^c = \vy_{ijkl}^c, \eta)] =
\eta\delta(\vy_{ijkl}^c, \vz )\quad \quad \quad ijkl \in {\mathscr D}
\end{equation}
where $\delta (\vy_{ijkl}^c, \vz ) = g(\vy_{ijkl}^c \cup\vy_{ijkl}^+, \vz) - g(\vy_{ijkl}^c \cup\vy_{ijkl}^-, \vz),$ the change
in $g(\vy, \vz)$ when $\vy_{ijkl}$ changes from $\vy_{ijkl}^{-}$ to $\vy_{ijkl}^{+}.$
The tetradic pseudo-likelihood for model (2) can then be defined as: 
\begin{equation}
\ell_{PT}(\eta;\vy) \equiv
\eta{\sum_{ijkl\in{\mathscr D}}{\delta (\vy_{ijkl}^c, \vz) \mathbb{I}(\vy_{ijkl}=\vy_{ijkl}^{+})}}
- \sum_{ijkl\in{\mathscr D}}\log\left[1+\exp(\eta\delta (\vy_{ijkl}^c, \vz))\right].
\label{eq:pltetradic}
\end{equation}
As $\vert{\mathscr D}\vert$ is large, we take a large random sample of 
them ($N=100000$) and use the sample mean to approximate (\ref{eq:pltetradic}). 
This procedure is
implemented in the {\tt statnet} {\tt R} package \citep{statnet}.

While the MPLE is know to be inferior to the MLE for dyadic dependence
models \citep{vanduijngilehan09} it is equivalent to the MLE for some dyadic
independence models. For the model (2) 
the network statistic is weakly dependent on the set of networks with the given degree and infection
sequences. Hence the maximum tetradic pseudo-likelihood (MTPLE) might be expected to perform well for this
model. This does seem to be the case for the models considered in this paper.
In simulations (not shown here) as it appears to be indistinguishable from
the MLE (where the later is computed by a computationally expensive MCMC
procedure). The advantages of the tetradic MPLE are that it is
computationally stable and fast while being numerically 
indistinguishable from the MCMC-MLE.  For these reasons we use it in all
simulations in this paper.

This estimator could be improved by adding hexadic
configurations to the pseudo-likelihood. These are necessary for sampling
algorithms to cover the full network space \citep{rao96}. However they also
lead to more complex algorithms and will be considered in other work.
\addcontentsline{toc}{section}{References}
\bibliographystyle{JASA_manu}
\bibliography{networksjasa}